\documentclass[11pt]{article} 
\usepackage{jheppub}

\usepackage{amsmath,amsfonts,amsbsy,amssymb,array,accents,dsfont}
\usepackage{enumerate}
\usepackage[shortlabels]{enumitem}
\usepackage{graphicx} 
\usepackage{subcaption} 
\usepackage{upgreek}
\usepackage{bm}
\usepackage{slashed}

\let\includefigures=\iftrue
\let\useblackboard==\iftrue
\newfam\black

\usepackage[dvipsames]{xcolor}

\definecolor{myblue}{RGB}{85,130,255}
\definecolor{myred}{RGB}{200, 45, 40}

\PassOptionsToPackage{ 
     colorlinks=true,
     linkcolor=myBlue,
     urlcolor=blue,
     filecolor=black,
     citecolor=myRed,
}{hyperref}

\usepackage{xparse}
\usepackage{xspace}

\NewDocumentCommand\eqn{om}{%
  \IfNoValueTF{#1}
     {\[ #2 \]}
     {\begin{equation}\label{#1} #2  \end{equation} \expandafter\newcommand\csname #1\endcsname{\eqref{#1}\xspace}\ignorespaces}
}
\NewDocumentCommand\eqna{om}{%
  \IfNoValueTF{#1}
    {\begin{align*} #2 \end{align*}}
    {\begin{equation}\label{#1}\begin{split} #2  \end{split}\end{equation} \expandafter\def\csname #1\endcsname{\eqref{#1}\xspace}\ignorespaces}
}

\newcommand{\rcite}{\cite}

\def\xx{{\bf x}}
\def\etaL{{\eta_L^{\vphantom{|}}}}
\def\etaR{{\eta_R^{\vphantom{|}}}}
\def\tauE{{\tau_{\!  \sst E}^{~}}}
\def\cst{{c_{\rm\sst ST}^{~}}}
\def\gsl{{g_\sl^{~}}}
\def\gsu{{g_\su^{~}}}

\def\alphab{{\boldsymbol\alpha}}

\def\sl{\text{sl}}
\def\su{\text{su}}

\def\vareps{\varepsilon}

\def\sltwo{\ensuremath{SU(1,1)}}
\def\sltwor{\ensuremath{SL(2,\bR)}}
\def\sltwoc{\ensuremath{SL(2,\bC)}}

\def\sutwo{{SU(2)}}

\def\uone{U(1)}

\def\hthree{{\bH_3^+}}

\def\tight#1{\! #1 \!}  

\def\({\left(}
\def\){\right)}
\def\[{\left[}
\def\]{\right]}

\def\ie{{i.e.}}
\def\eg{{e.g.}}

\def\etc{{etc}}

\def\lstr{\ell_{\textit{s}}}

\def\gstrsq	{g_{\textit s}^{2}}

\def\nfive{{n_5}}

\def\flabel{{\sst (m)}}

\def\A{{\mathsf A}}
\def\B{{\mathsf B}}
\def\C{{\mathsf C}}
\def\F{{\mathsf F}}
\def\H{{\mathsf H}}
\def\K{{\mathsf K}}

\def\sfM{{\mathsf M}}

\def\sfa{{\mathsf a}}

\def\sfk{{\mathsf k}}
\def\sfm{{\mathsf m}}
\def\sfn{{\mathsf n}}
\def\sfp{{\mathsf p}}

\DeclareMathSymbol{\medhatsym}{\mathord}{largesymbols}{"62} 

\DeclareMathSymbol{\medtildesym}{\mathord}{largesymbols}{"65}

\makeatletter
\newcommand*\rel@kern[1]{\kern#1\dimexpr\macc@kerna}
\newcommand*\widebar[1]{%
  \begingroup
  \def\mathaccent##1##2{%
    \rel@kern{0.8}%
    \overline{\rel@kern{-0.8}\macc@nucleus\rel@kern{0.2}}%
    \rel@kern{-0.2}%
  }%
  \macc@depth\@ne
  \let\math@bgroup\@empty \let\math@egroup\macc@set@skewchar
  \mathsurround\z@ \frozen@everymath{\mathgroup\macc@group\relax}%
  \macc@set@skewchar\relax
  \let\mathaccentV\macc@nested@a
  \macc@nested@a\relax111{#1}%
  \endgroup
}
\makeatother

\def\Ry{R_y}

\def\half{\frac12}

\def\tr{{\rm Tr}}

\def\One{{\hbox{1\kern-1mm l}}}

\def\barray{\begin{array}}
\def\earray{\end{array}}
\def\be{\begin{equation}}
\def\ee{\end{equation}}
\def\bea{\begin{eqnarray}}
\def\eea{\end{eqnarray}}
\def\bal{\begin{align}}
\def\eal{\end{align}}
\def\nn{\nonumber}

\newcommand{\bC}{{\mathbb C}}

\newcommand{\bH}{{\mathbb H}}

\newcommand{\bR}{{\mathbb R}}
\newcommand{\bS}{{\mathbb S}}
\newcommand{\bT}{{\mathbb T}}

\newcommand{\bZ}{{\mathbb Z}}

\definecolor{cardinal}{rgb}{0.6,0,0}
\definecolor{darkgreen}{rgb}{0,0.4,0}
\definecolor{green}{rgb}{0,0.4,0}
\definecolor{golden}{rgb}{0.92, 0.7, 0}
\definecolor{midnight}{rgb}{0, 0, 0.5}
\definecolor{darkblue}{rgb}{0, 0, 0.7}

\numberwithin{equation}{section}

\mathchardef\mhyphen="2D

 \def\cB{\mathcal {B}} 
 \def\cE{\mathcal {E}} 
\def\cG{\mathcal {G}} \def\cH{\mathcal {H}} 
\def\cJ{\mathcal {J}} \def\cK{\mathcal {K}} \def\cL{\mathcal {L}}
\def\cM{\mathcal {M}}  \def\cO{\mathcal {O}}
  \def\cR{\mathcal {R}}
  
\def\cV{\mathcal {V}}  
\def\cY{\mathcal {Y}}

\def\one{{\hbox{\kern+.5mm 1\kern-.8mm l}}}
\def\zero{{\hbox{0\kern-1.5mm 0}}}

\newcommand{\ket}[1]{{\,| {#1} \rangle}}

\def\id{\textrm{id}}

\def\id{{1 \kern-.28em {\rm l}}}

\def\journal#1&#2(#3){\unskip, \sl #1\ \bf #2 \rm(19#3) }
\def\andjournal#1&#2(#3){\sl #1~\bf #2 \rm (19#3) }

\def\ie{{\it i.e.}}
\def\eg{{\it e.g.}}

\def\etc{{\it etc}}

\def\sst{\scriptscriptstyle}

\def\half{\frac12}

\def\ket#1{|#1\rangle}

\def\vev#1{\langle#1\rangle}

\def\One{{1\hskip -3pt {\rm l}}}

\catcode`\@=11
\def\slash#1{\mathord{\mathpalette\c@ncel{#1}}}
\def\vareps{\varepsilon}
\def\underrel#1\over#2{\mathrel{\mathop{\kern\z@#1}\limits_{#2}}}

\catcode`\@=12

\def\ket#1{\left| #1\right\rangle}
\def\vev#1{\left\langle #1 \right\rangle}
\def\det{{\rm det}}

\def\det{{\rm det}}
\def\exp{{\rm exp}}

\def\ie{{\it i.e.}}
\def\eg{{\it e.g.}}

\title{
{
$\bf AdS_3$ Orbifolds, BTZ Black Holes, and Holography
}}

\author{
Emil J. Martinec
}

\affiliation{
\vskip 0.01cm
Kadanoff Center for Theoretical Physics, Enrico Fermi Institute, and Department of Physics\\ 
University of Chicago,
5640 S. Ellis Ave.,
Chicago IL 60637\\ 
}

\emailAdd{%
e-martinec@uchicago.edu}

\abstract{%
Conical defects of the form $(AdS_3\times \bS^3)/\bZ_\sfk$ have an exact orbifold description in worldsheet string theory, which we derive from their known presentation as gauged Wess-Zumino-Witten models.  
The configuration of strings and fivebranes sourcing this geometry is well-understood, as is the correspondence to states/operators in the dual $\it CFT_2$.  
One can analytically continue the construction to Euclidean $AdS_3$ (\ie\ the hyperbolic ball $\hthree$) and consider the orbifold by any infinite discrete (Kleinian) group generated by a set of elliptic elements $\gamma_i\in \sltwoc$, $\gamma_i^{\sfk_i}=\One$, $i=1,...,K$.  The resulting geometry consists of multiple conical defects traveling along geodesics in $\hthree$, and provides a semiclassical bulk description of correlation functions in the dual CFT involving the corresponding defect operators, which is nonperturbatively exact in $\alpha'$.  The Lorentzian continuation of these geometries describes a collection of defects colliding to make a BTZ black hole.  We comment on a recent proposal to use such correlators to prepare a basis of black hole microstates, and elaborate on a picture of black hole formation and evaporation in terms of the underlying brane dynamics in the bulk.
}

\begin{document}
\hypersetup{pageanchor=false}
\begin{titlepage}
\maketitle
\thispagestyle{empty}
\end{titlepage}
\hypersetup{pageanchor=true}
\pagenumbering{arabic}

\thispagestyle{empty}







\section{Introduction and summary} 
\label{sec:intro}

The $AdS_3/CFT_2$ correspondence has proven to be one of the richest examples of gauge/gravity duality.  On the one hand, two-dimensional conformal symmetry is especially powerful, providing a remarkable degree of analytic control over the conformal field theory.
On the other, there is a large catalogue of exact, asymptotically $AdS_3$ supergravity solutions (see~\rcite{Mathur:2005zp,Bena:2007kg,Shigemori:2020yuo} for reviews).  In addition, there are exactly solvable worldsheet theories describing perturbative string theory around global $AdS_3$~\rcite{Giveon:1998ns,Kutasov:1999xu,Maldacena:2000hw,Maldacena:2001km}, as well as certain conical defect geometries~\rcite{Martinec:2017ztd}, and the BTZ black hole~\rcite{Natsuume:1996ij,Maldacena:2000kv,Hemming:2001we,Hemming:2001we,Troost:2002wk,Hemming:2002kd}.

The $(AdS_3\times\bS^3)/\bZ_\sfk$ conical defect geometries, whose worldsheet description was elaborated in~\rcite{Martinec:2017ztd,Martinec:2018nco,Martinec:2019wzw,Martinec:2020gkv,Bufalini:2021ndn,Martinec:2022okx}, provide rare examples where one has a fully stringy description of bulk states far from the vacuum as well as their holographic map to the spacetime CFT.  These backgrounds arise as particular geometries sourced by bound states of fundamental (F1) strings and NS5-branes in a decoupling limit.  There is a precise map between the geometry and the fivebranes' source configuration which is in turn determined by the particular condensate of fundamental strings they carry~\rcite{Lunin:2002bj,Kanitscheider:2006zf,Kanitscheider:2007wq}.  This map is reflected in the structure and properties of worldsheet vertex operators~\rcite{Martinec:2020gkv,Martinec:2022okx}, in that winding string vertex operators act to perturb the background string condensate.
Our goal here is to develop and generalize the worldsheet theory of these orbifold geometries, and their connection to gauge-gravity duality.

The worldsheet theory was initially constructed as a gauged WZW model 
\be
\label{GmodH}
\frac{\cG}{\cH} = \frac{\sltwo\times\sutwo \times \bR_t\times \bS^1_y \times \bT^4}{U(1)_L\times U(1)_R}
\ee
where $\cH$ embeds in $\cG$ as a pair of left and right null isometries.  These worldsheet CFT's in fact describe more general asymptotically linear dilaton geometries rather than asymptotically $AdS_3$, but have a modulus (the radius $R_y$ of $\bS^1_y$) for which $AdS_3$ asymptotics arises in the $R_y\to\infty$ limit.

It is essential for the purposes of the present work to have a direct worldsheet description of the conical defects $(AdS_3\times\bS^3)/\bZ_\sfk$ as orbifold CFT's.  To this end, in section~\ref{sec:nullgauging} we show how the physical state spectrum of the theory~\eqref{GmodH} becomes that of a standard orbifold in the $R_y\to\infty$ limit, and exhibit the orbifold identification.  We also show that the linear dilaton extension can be thought of as a current-current deformation of the orbifold CFT, where the current is none other than the $\cK=AdS_3\times\bS^3$ part of the gauge currents $\cJ,\bar \cJ$ of the $\cG/\cH$ gauging.  In hindsight, this result appears to be a special case of a general relation between gauged WZW models $(\cK\times\cH)/\cH$, current-current deformations of WZW models on $\cK$, and their orbifolds~\rcite{Forste:2003km}.

The description of a class of $AdS_3$ conical defects as string theory orbifolds paves the way for a series of generalizations that occupy the remainder of our analysis.  With the standard orientation of the orbifold group, the defect sits statically at the origin of $AdS_3$ and (as we will see below) along a circle in $\bS^3$.  But one can use the $\sltwo_L\times\sltwo_R$ isometries of $AdS_3$ to conjugate the orbifold identification into a different but equivalent $\bZ_\sfk$ action, which in general describes boosted and orbiting defects.

\begin{figure}[ht]
\centering
  \begin{subfigure}[b]{0.25\textwidth}
  \hskip 0cm
    \includegraphics[width=\textwidth]{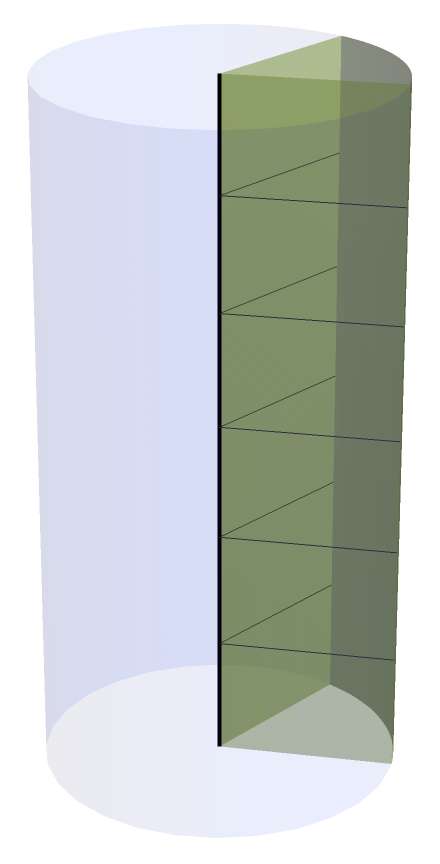}
    \caption{ }
    \label{fig:Defect}
  \end{subfigure}
\qquad
  \begin{subfigure}[b]{0.25\textwidth}
      \hskip 0cm
    \includegraphics[width=\textwidth]{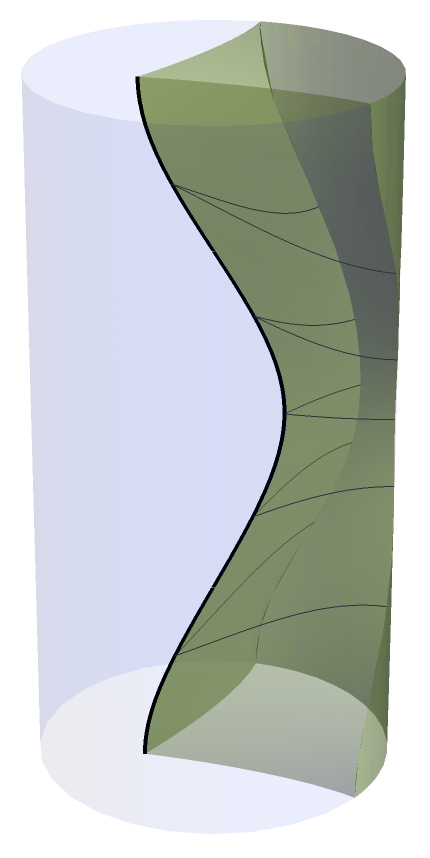}
    \caption{ }
    \label{fig:DefectBoost}
  \end{subfigure}
\qquad
  \begin{subfigure}[b]{0.25\textwidth}
      \hskip 0cm
    \includegraphics[width=\textwidth]{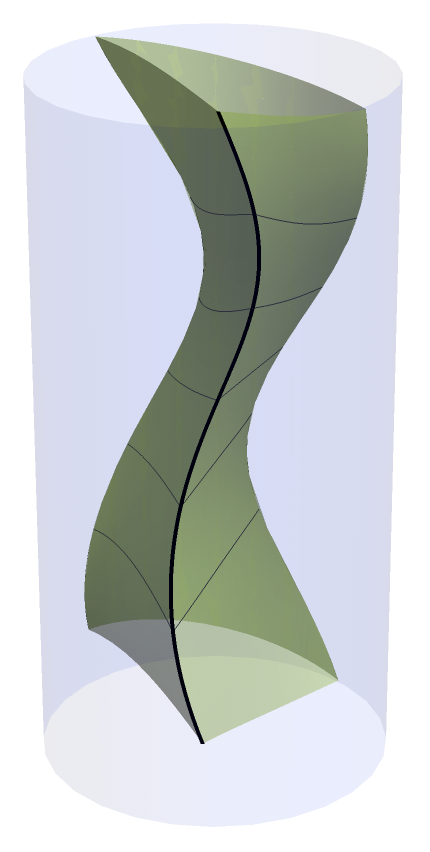}
    \caption{ }
    \label{fig:DefectRot}
  \end{subfigure}
\caption{\it 
(a) A static conical defect sits in the center of $AdS_3$ and removes a wedge from the geometry, with the two sides of the wedge identified by a rotation around the origin (in this case by an angle $2\pi/6$).  
(b) A radially boosted defect oscillates about the center of $AdS_3$, and removes a boosted wedge, with the sides identified by the same rotation conjugated by the boost.  
(c) A boosted and rotating defect.  
The locus of points being identified on several timeslices is indicated in each figure.
(Note that the defect angle here is simply for illustrative purposes, and does not correspond to the defect angle $2\pi(1-1/\sfk)$ of a $\bZ_\sfk$ orbifold geometry.)
}
\label{fig:ptcles}
\end{figure}

The standard defect geometries (an example of which is shown in figure~\ref{fig:Defect}) are dual to particular heavy states $\ket{\Psi_\sfk}$ in the spacetime CFT; and to each such state is associated the operator $\cO_\sfk(z,\bar z)$ that creates it from the CFT vacuum~$\ket\Omega$ via
\be
\ket{\Psi_\sfk} = \cO_\sfk(0,0)\ket{\Omega} ~.
\ee
One can track the 1/2-BPS operators across the moduli space of the spacetime CFT to weak coupling, where the spacetime CFT is a symmetric product orbifold $(\bT^4)^N/S_N$.
The CFT operators $\cO_\sfk$ are known symmetric group orbifold twist operators~\rcite{Lunin:2002bj}; in fact the holographic map for all 1/2-BPS states is understood~\rcite{Kanitscheider:2006zf,Kanitscheider:2007wq}.

The conjugation operation
\be
\cO_\sfk(z,\bar z) = e^{z \cL_{-1}+\bar z \bar \cL_{-1}} \, \cO_\sfk(0,0)\, e^{-(z \cL_{-1}+\bar z \bar \cL_{-1})}
\ee
is the Euclidean CFT version of a global $\sltwo_L\times\sltwo_R$ isometry of $AdS_3$~-- on the conformal boundary of Euclidean $AdS_3$, these symmetries translate the operator to a more general position.  The translated operators are dual (after analytic continuation to Lorentz signature) to the bulk conical defect oscillating radially about the center of $AdS_3$ as in figures~\ref{fig:DefectBoost}, or boosted and rotating as in figure~\ref{fig:DefectRot}.

\begin{figure}[ht]
\centering
  \begin{subfigure}[b]{0.25\textwidth}
  \hskip 0cm
    \includegraphics[width=\textwidth]{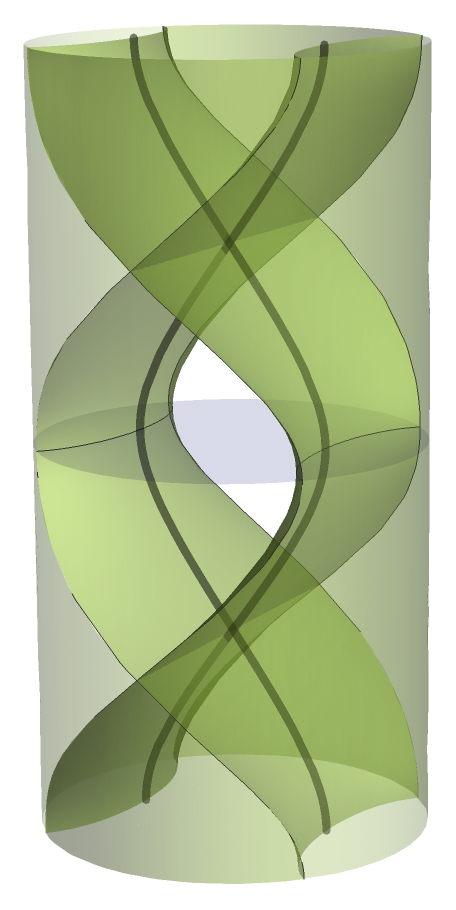}
    \caption{ }
    \label{fig:Z2-BTZ}
  \end{subfigure}
\qquad\qquad
  \begin{subfigure}[b]{0.28\textwidth}
  \vskip -2cm
    \includegraphics[width=\textwidth]{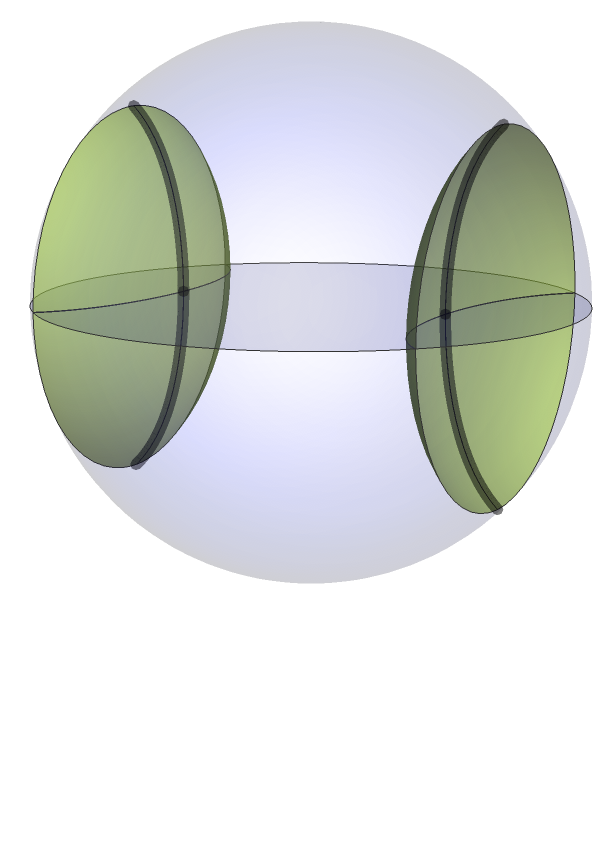}
    \caption{ }
    \label{fig:EucZ2-3D}
  \end{subfigure}
\caption{\it 
a) The collision of two $\bZ_2$ defects, obtained by identifying global $AdS_3$ by separate $\pi$ rotations about oppositely boosted geodesics (the thick black lines), makes a BTZ black hole.  The surfaces being identified on either side of the geodesic are shaded green, as is the excluded region of the conformal boundary.  The lightly shaded horizontal disk indicates the surface of time reflection symmetry, on which both defects are instantaneously at rest.  The past and future BTZ singularities are the locus where the sheets being identified collide; the unshaded region between the sheets is the physical BTZ geometry (the regions above and below the singularity contain closed timelike curves).  Note that due to the identifications, there is only one connected asymptotic region.  The future (past) horizons are the past (future) light cones of the points where the singularity meets the conformal boundary.  \\
$~~~~$
b) Euclidean continuation of the two-defect geometry has the two $\bZ_2$ defects traveling geodesics in $\hthree$, the Euclidean continuation of $AdS_3$.  The two spherical shells on either side of each geodesic are identified.  The Lorentzian and Euclidean geometries agree on the surface of time reflection symmetry (the grey Poincar\'e disk).  One can thus think of preparing the Lorentzian geometry on this surface by a Euclidean evolution from the vacuum with operator insertions, using the lower half of the Euclidean geometry \`a la Hartle-Hawking, to make defects instantaneously at rest as an initial condition for Lorentzian evolution.
}
\label{fig:twoZ2}
\end{figure}

Having moved the defect off to one side in $AdS_3$, one is free to introduce another; and another, and so on.  The collision of such defects has been studied in~\rcite{Matschull:1998rv,Holst:1999tc,Birmingham:1999yt,Krasnov:2002rn,DeDeo:2002yg,Brill:2007zq,Lindgren:2015fum}; we review and extend these results in section~\ref{sec:conjugation}.  With sufficient aggregate mass and/or radial momentum, the collision of the defects forms a BTZ black hole.   
As we will see below, the collision of $\bZ_\sfk$ orbifold defects always makes a black hole.  We will examine the geometries arising from the global orbifolds describing colliding $\bZ_\sfk$ defects, and exhibit some of their properties, first in section~\ref{sec:radialboost} for multiple defects moving purely radially as in figure~\ref{fig:DefectBoost}, and then in section~\ref{sec:rotating} for multiple defects orbiting as in figure~\ref{fig:DefectRot}.  The example of two $\bZ_2$ defects radially boosted in opposite directions is shown in figure~\ref{fig:Z2-BTZ}.  

The presence of the BTZ singularity causes perturbative string theory to break down.  This was seen in the similar situation of Lorentzian orbifolds of flat spacetime string theory~\rcite{Liu:2002kb,Horowitz:2002mw}.
Indeed, we shouldn't expect perturbative string theory to resolve all the issues arising from black holes in quantum gravity.  But one thing we can do is rotate the theory to Euclidean signature where the black hole singularity, as well as other pathologies such as closed timelike curves, are absent.  Geometrically, Euclidean BTZ is $\hthree/\bZ$; the worldsheet theory has been studied in~\rcite{Natsuume:1996ij,Maldacena:2000kv,Hemming:2001we,Troost:2002wk,Hemming:2002kd,Rangamani:2007fz,Berkooz:2007fe,Lin:2007gi,Ashok:2020dnc,Nippanikar:2021skr,Ashok:2021ffx,Ashok:2022vdz,Kaundinya:2023xoi}.  Here we extend the discussion to orbifolds $\hthree/\Gamma$ where $\Gamma\in\sltwoc$ is a particular class of discrete (Kleinian) group generated by elliptic transformations $\gamma_i\in\sltwoc$, $\gamma_i^{\sfk_i}=\One$ (see also~\rcite{Krasnov:2001va,Chandra:2023dgq}).%
\footnote{The discussion can of course be extended to arbitrary Kleinian groups, providing a string theory realization of the ideas in~\rcite{Maloney:2015ina}.}

We thus examine the continuation of Lorentzian $AdS_3$ to its Euclidean counterpart $\hthree$ in section~\ref{sec:euclidean}.  We first show how the $\sltwo_L\times\sltwo_R$ isometries of $AdS_3$ are related to the $\sltwoc$ isometry of $\hthree$ under analytic continuation, and then exhibit the corresponding defect geometries.  For radially boosted (\ie\ non-orbiting) defects, the geometry can be chosen to have a surface of time reflection symmetry where all the orbifold defects are momentarily at rest in $AdS_3$.  This spacelike hypersurface has the geometry of the Poincar\'e disk $\bH_2$ which is also a surface of time reflection symmetry of $\hthree$ under the natural Wick rotation of the global time in $AdS_3$; furthermore, $\Gamma$ acts as a set of identifications {\it within} this $\bH_2$.  The two geometries share this common hypersurface $\bH_2/\Gamma$, and one can think of passing from one to the other along this hypersurface~-- as one does for instance in the Schwinger-Keldysh formalism~\rcite{Haehl:2016pec,Haehl:2016uah,deBoer:2018qqm}, or the Hartle-Hawking method of state preparation via a period of Euclidean evolution~\rcite{Hartle:1983ai}.  For the case of orbiting defects, there is no time reflection symmetry, but we can still say how the full spacetimes are related.

Figure~\ref{fig:EucZ2-3D} depicts the analytic continuation to Euclidean signature of the Lorentzian geometry sourced by two $\bZ_2$ defects shown in figure~\ref{fig:Z2-BTZ}.  We are then solidly grounded in the arena of standard, well-behaved string theory orbifolds, and the construction of perturbative string theory is in principle straightforward.
The Euclidean defect geometries again have the defects traveling geodesics in $\hthree$.  These are circles which orthogonally intersect the $\bS^2$ conformal boundary of $\hthree$.  The surfaces being identified to make the defect are spherical wedges identified under a $\bZ_\sfk$ elliptic element of $\sltwoc$, the isometry group of $\hthree$.

The worldsheet theory in the presence of a single defect provides a route to computing (order by order in an expansion in powers of the string coupling) the spacetime CFT correlation functions involving several ``light'' operators in the presence of two ``heavy'' operators.  Here the light operators are those that introduce and remove perturbative strings, and therefore have spacetime conformal dimension $h,\bar h$ much less than the central charge $\cst$ of the spacetime CFT.  The two heavy operators have conformal dimension $h,\bar h\sim O(\cst)$, and serve to introduce and remove the defect~\rcite{Galliani:2017jlg,Bombini:2017sge,Bufalini:2022wzu}.  Correlation functions of string theory on $\hthree$ have been investigated in~\rcite{Kutasov:1999xu,Maldacena:2001km,Teschner:1999ug,Teschner:2001gi,Ponsot:2002cp,Hikida:2007tq,Dei:2021xgh,Dei:2021yom,Dei:2022pkr,Bufalini:2022toj}, and these particular single-defect correlators are closely related due to the orbifold structure~\rcite{Bufalini:2022wzu}.

The more general orbifolds of $\hthree/\Gamma$ describe correlators of light operators in the presence of multiple heavy operators.  Such correlators have been studied from the spacetime CFT point of view for instance in~\rcite{Raeymaekers:2022sbu,Chandra:2023dgq}, where they were used to investigate properties of the holographic map, and in particular whether observables are inside or outside the apparent horizon at the surface of time reflection symmetry.  The construction here allows the possibility to put these ideas to the test in string theory.  For instance, the worldsheet two-point function should be the sort of diagnostic of the holographic embedding envisaged by~\rcite{Chandra:2023dgq}, and should provide a useful check of this proposal.


The string worldsheet correlators for the non-abelian orbifolds $(\hthree\times\bS^3)/\Gamma$ thus compute the semi-classical approximation to spacetime CFT correlation functions involving multiple heavy and light operator insertions.  Their evaluation is a rather involved technical exercise, which we will leave to future work; we can however at least characterize them.  For instance, in the effective supergravity theory, two-point functions of light operators in the orbifold background will be given by Green functions for the appropriate Laplace operator on $\hthree/\Gamma$.  The untwisted sector consists of solutions of the wave equation on $\hthree$ that are invariant under $\Gamma$, which can formally be constructed by the method of images (regulating and renormalizing the sums involved); we will touch on this issue briefly below in section~\ref{sec:correlators}.  In addition, each defect will have its own twisted sectors which describe wound strings pinned to the defect.  There will also be a continuum of ``long string'' states~\rcite{Maldacena:2000hw} that have plane wave behavior out near the conformal boundary and thus surround all of the defects.  The latter are related to the Hawking radiation of wound strings discussed recently in~\rcite{Martinec:2023plo,Martinec:2023iaf}.

As in the case of a single defect, one expects the winding string vertex operators on $(\hthree\times\bS^3)/\Gamma$ to perturb the string condensate carried by the fivebranes, just as they do for a single defect.  Non-abelian fivebrane dynamics underlies that of BTZ black holes in string theory and is governed by {\it little string theory}~\rcite{Dijkgraaf:1997nb,Dijkgraaf:1997ku,Maldacena:1996ya,Seiberg:1997zk}, with the black hole microstates being accounted for by the Hagedorn entropy of the little string~\rcite{Maldacena:1996ya,Martinec:2019wzw}; in perturbative regimes, fundamental strings are composites of the little string whose condensation coherently deforms the fivebrane state~\rcite{Martinec:2020gkv,Martinec:2022okx}.  There have been suggestions~\rcite{McGreevy:2005ci,Horowitz:2006mr,Berkooz:2007fe} that a perturbative string winding condensate is involved in the resolution of the BTZ singularity; the effect of such a condensate would however naively be expected to be confined to the region where the proper size of such strings is of order the string scale, which is quite close to the singularity.  But one needs effects that persist out to the horizon scale which is arbitrarily larger than the string scale; it is hard to see how effects involving perturbative strings near the singularity could affect the black hole interior sufficiently to resolve the fundamental tension between causality, locality and unitarity posed by the Hawking process.  As observed in~\rcite{Martinec:2019wzw}, however, the little string tension is such that it is always at its correspondence point, which would mean that the quantum wavefunction of the little string could extend to the horizon scale, in accordance with the fuzzball paradigm (see~\rcite{Mathur:2005zp,Bena:2022rna} for reviews).  One may hope to gain enough control over the worldsheet construction to begin to see the outlines of this scenario take shape, as the conical defect collisions described here start to form a BTZ black hole.  Indeed, we will take some first steps in this direction in section~\ref{sec:correlators} when we consider two $\bZ_2$ defects in the extremal limit, where their OPE is BPS protected and produces a Hagedorn little string gas.

The wide variety of $(\hthree\times\bS^3)/\Gamma$ orbifold geometries described here all lead to pure state BTZ black holes under Lorentzian continuation.  This method of preparing black hole states has been discussed recently in~\rcite{Chandra:2022fwi,Chandra:2023dgq}; and also in~\rcite{Balasubramanian:2022gmo,Balasubramanian:2022lnw}, where it was suggested that such states could form a basis of black hole microstates.  The analysis in these works seeks to be quite general, applying to any holographic duality (and especially for $\it AdS_3/CFT_2$).  It is however quite useful to drill down on a specific instance of the duality where we know a lot about its embedding in string theory, so that we can be sure we are not presupposing properties of the duality for which there are no concrete realizations, or objects whose holographic map is poorly understood.  Furthermore, a major unresolved issue in these constructions is what happens to the state under Lorentzian evolution once we have prepared it.  The string theoretic construction we provide here yields some hints in that regard, given our extensive knowledge of the holographic map and in particular its stringy properties.  We will elaborate on these issues in section~\ref{sec:discussion}.


\section{Null-gauged WZW models and orbifolds of $\bf AdS_3$}
\label{sec:nullgauging}

We approach the description of $(AdS_3\times \bS^3)/\bZ_\sfk$ orbifolds via a detour into the null-gauged WZW models studied in~\rcite{Martinec:2017ztd,Martinec:2018nco,Martinec:2019wzw,Martinec:2020gkv,Bufalini:2021ndn,Martinec:2022okx}.  These models have linear dilaton rather than $AdS$ asymptotics, and so do not directly describe the $AdS_3$ conical defects; however they have a modulus $R_y$ such that the $AdS$ orbifold geometry is recovered in the limit $R_y\to\infty$.  Our strategy will be to deduce the orbifold action on $\sltwo\times \sutwo $ from the $AdS$ limit of the null-gauged model~\eqref{GmodH}, and then relate the general null-gauged model to a marginal current-current deformation of this orbifold.  For those less interested in the details of the worldsheet theory, this section can be skipped on a first reading.

These isometries are generated by left and right null currents
\begin{align}
\begin{split}
\cJ &= J^3_\sl + \ell_2 J^3_\su + \ell_3 P_t +\ell_4 P_{y}
\\[.3cm]
\bar\cJ &= \bar J^3_\sl +  r_2 \bar J^3_\su + r_3 \bar P_t + r_4 \bar P_{y}  ~,
\end{split}
\end{align}
where both $\sltwo$ and $\sutwo$ current algebras are at level $n_5$.
Integrating out the gauge fields leads to an effective sigma model geometry with the linear dilaton asymptotics of a decoupled fivebrane throat, while a suitable choice of null vector coefficients $\ell_i,r_i$ 
\begin{align}
\begin{split}
\label{nullvecs}
\ell_2 &= -(\sfn +\sfm) \in 2\bZ+1 
~~,~~~~
r_2 = \sfn -\sfm\in 2\bZ+1
~~,~~~~
\sfm,\sfn \in \bZ
\\[.2cm]
\ell_4 &= \frac{\sfp}{R_y}+\sfk R_y
~~,~~~~
r_4 = \frac{\sfp}{R_y} - \sfk R_y
~~,~~~~ \sfp,\sfk\in\bZ
\\[.2cm]
\ell_3 &= r_3 = - \sqrt{\sfk^2 R_y^2 +\frac{\sfp^2}{R_y^2} + n_5(\sfm^2+\sfn^2-1)} ~,
\end{split}
\end{align}
subject to the constraint
\be
\label{kpmn}
\sfk \sfp = -n_5 \sfm \sfn ~,
\ee
leads to IR asymptotics of the form $(AdS_3\times\bS^3)/\bZ_\sfk$ and UV asymptotics given by the linear dilaton throat of the $n_5$ fivebranes~\rcite{Martinec:2017ztd,Martinec:2018nco,Bufalini:2021ndn}.  

When $\sfp\tight=\sfn\tight=0$. $\sfm\tight=1$, the backgrounds are the 1/2-BPS conical defect geometries of Lunin and Mathur~\rcite{Lunin:2001fv,Lunin:2001jy}; generalizing to $\sfm-\sfn=1$ (while satisfying~\eqref{kpmn}) realizes the 1/4-BPS backgrounds of~\rcite{Giusto:2004id,Giusto:2012yz}; the generic solution breaks all spacetime supersymmetries and describes the ``JMaRT'' family of solutions of~\rcite{Jejjala:2005yu,Chakrabarty:2015foa}.  

The crossover between the UV linear dilaton regime and the IR $AdS_3$ regime is controlled by the modulus $R_y$.  Heuristically, the null gauging relates the azimuthal direction of $\sltwo=AdS_3$ to the circle $\bS^1_y$.  The proper size of the remaining gauge invariant circle grows exponentially with radius as it does in $AdS_3$ until it saturates at the scale $R_y$ and the geometry rolls over into the linear dilaton background.  

Sending $R_y\to\infty$ decouples the linear dilaton regime, so that the entire geometry takes the form $(AdS_3\times \bS^3)/\bZ_\sfk$.  In this limit, the gauge current becomes dominated by the contributions from $\bR_t\times \bS^1_y$, and the gauge orbits largely lie along these directions.  It is then convenient to fix the gauge $t=y=0$, which leaves behind a residual discrete symmetry that acts to quotient of $\sltwo\times\sutwo = AdS_3 \tight\times \bS^3$.
The structure of the gauge orbits was described in detail in~\rcite{Martinec:2018nco,Martinec:2019wzw}.

\subsection{1/2-BPS embeddings $\cH\hookrightarrow\cG$}
\label{sec:STembed}

Consider to begin with the 1/2-BPS geometries; the general element $(e^{i\alpha},e^{i\beta})\in U(1)_L\times U(1)_R$ acts on $\cG$ as
\be
\label{Horbit}
\Big( g_\sl, g_\su, t, y, \tilde y \Big) \mapsto
\Big( e^{i\alpha\sigma_3} g_\sl e^{i\beta\sigma_3}, 
e^{-i\alpha \sigma_3} g_\su e^{-i\beta\sigma_3}, 
t-\sfk R_y(\alpha+\beta),
y+\sfk R_y(\alpha-\beta),
\tilde y-\sfk R_y(\alpha+\beta) \Big)
\ee
where $\tilde y$ is the coordinate that parametrizes the circle T-dual to $\bS^1_y$.  The orbits of the axial gauge motion $\alpha=-\beta$ are compact with period $2\pi$.  We use this gauge motion to fix $y$, but this only uses up a fraction $1/\sfk$ of the gauge orbit; there is a residual $\bZ_\sfk$ discrete remnant, which implements a discrete quotient on $\sltwo\times\sutwo$.

The spectrum of $\cG$ consists of current algebra descendants built on the highest weight vertex operators
\be
\label{comvert}
\Phi^{(w)}_{j,m,\bar m} \, \Psi^{(w',\bar w')}_{j',m',\bar m'} \, \exp\big[ {-iEt+i(n_y /R_y)y+i(w_y R_y)\tilde y} \big]~,
\ee
where $\Phi^{(w)}_{j,m,\bar m}$ is a primary in the spectral flow sector $w$ of supersymmetric $\sltwo$ current algebra, and similarly $\Psi^{(w',\bar w')}_{j',m',\bar m'}$ is a primary in the spectral flow sector $(w',\bar w')$ of the supersymmetric $\sutwo$ theory.  The axial null constraint $\cJ-\bar\cJ=0$ imposes
\be
\big( m-\bar m\big) + (M-\bar M) - \Big( m'-\bar m'+\frac{n_5}{2}(w'-\bar w') + (M'-\bar M') \Big) = k n_y
\ee
where $M-\bar M$ is the net $J^3_\sl-\bar J^3_\sl$ charge of the $\sltwo$ descendants (and similarly for $M'-\bar M'$ and $\sutwo$).  The effect of the residual discrete gauge symmetry is to project the momentum along the $\sltwo\times \sutwo$ component of the axial gauge orbit onto an integer multiple of $\sfk$, \ie\ just what would expect for a $\bZ_\sfk$ orbifold projection.  The result is an orbifold if the rest of the BRST invariant spectrum accounts for the twisted sectors of the orbifold.  Let us see how this arises in the $R_y\to\infty$ limit.

The left/right $y$ momenta are
\be
P_{y} = \frac{n_y}{R_y}+w_yR_y
~~,~~~~
\bar P_y = \frac{n_y}{R_y} - w_y R_y ~.
\ee 
We then write
\be
E = w_y R_y + \frac{\vareps}{R_y}
\ee
since the string winding accounts for the bulk of the energy in the limit of large $R_y$; since the constraints are solved order by order in $R_y$, $\vareps$ is independent of $R_y$ at leading order.  The null constraints in this limit set
\begin{align}
\begin{split}
\label{largeRnull}
\sfk(\vareps-n_y) &= 2m+n_5 w-(2m'+n_5w') + M-M'
\\[.2cm]
\sfk(\vareps+n_y) &= 2\bar m+n_5 w-(2\bar m'+n_5\bar w') + \bar M-\bar M' ~,
\end{split}
\end{align}
while the Virasoro constraints in this limit impose
\begin{align}
\begin{split}
\label{largeRvir}
\half &= \frac{-j(j-1)+j'(j'+1)}{n_5} - (m+M)w+(m'+M')w'+\frac{n_5}4\big(w^2-(w')^2\big) + w_y(\vareps - n_y) +N_L
\\[.3cm]
\half &= \frac{-j(j-1)+j'(j'+1)}{n_5} - (\bar m+\bar M)w+(\bar m'+\bar M')\bar w'+\frac{n_5}4\big(w^2-(\bar w')^2\big) + w_y(\vareps + n_y) +N_R
\end{split}
\end{align}
where $N_L,N_R$ are the oscillator excitation levels.  

Substituting the values of $\vareps\pm n_y$ from the null constraints into the Virasoro constraints, one sees that the effect of $w_y$ in the Virasoro constraints can be absorbed in a simultaneous shift 
\be
w\to w+ w_y/\sfk
~~,~~~~
w'\to w'+ w_y/\sfk
~~,~~~~
\bar w'\to \bar w'+ w_y/\sfk
\ee
which can be thought of as a simultaneous fractional spectral flow in both $\sltwo$ and $\sutwo$ (the quadratic terms in $w_y$ cancel between the two, leaving just the terms linear in $w_y$).  In other words, the spectrum of vertex operators with $t,y$ present is the same as that of $\cK=\sltwo\times\sutwo$ alone, with an orbifold projection on the untwisted sector and fractional spectral flow accounting for the twisted sectors.

Bosonizing the currents in terms of canonically normalized scalar fields $\cY_\sl,\cY_\su$
\be
\label{bosonize}
J^3_\sl = \sqrt{n_5}\, \partial\cY_\sl
~~,~~~~
\bar J^3_\sl = \sqrt{n_5}\, \bar \partial\bar \cY_\sl
~~,~~~~
J^3_\su = \sqrt{n_5}\, \partial\cY_\su
~~,~~~~
\bar J^3_\su = \sqrt{n_5}\, \bar \partial\bar \cY_\su  ~,
\ee
one can factor out the charge dependence from the vertex operators~\eqref{comvert} as
\begin{align}
\begin{split}
\label{vertexops}
\Phi^{(w)}_{j,m,\bar m} &= V^{(w)}_{j,m,\bar m}\, \exp\Big[\frac{2}{\sqrt{n_5}}\Big( m+\frac{n_5}{2} w \Big)\cY_\sl + \frac{2}{\sqrt{n_5}}\Big( \bar m+\frac{n_5}{2} w \Big)\bar \cY_\sl \Big]
\\[.3cm]
\Psi^{(w',\bar w')}_{j',m',\bar m'} &= \Lambda^{(w',\bar w')}_{j',m',\bar m'}\, \exp\Big[\frac{2}{\sqrt{n_5}}\Big( m'+\frac{n_5}{2} w' \Big)\cY_\su + \frac{2}{\sqrt{n_5}}\Big( \bar m'+\frac{n_5}{2} \bar w' \Big)\bar \cY_\su \Big]  ~.
\end{split}
\end{align}
The null gauged WZW is then equivalent in the $R_y\to\infty$ limit to a standard $\bZ_\sfk$ shift orbifold generated~by
\be
\label{shiftorb}
\delta\big(\cY_\sl-\bar \cY_\sl\big) = -\delta\big(\cY_\su - \bar\cY_\su\big) = \frac{2\pi}{\sfk}  ~.
\ee
The axial null constraint restricts the difference of the $\cY$ momenta $J^3_\sl+\bar J^3_\sl$ and $J^3_\su+\bar J^3_\su$ to be a multiple of $\sfk$, and the effect of the vector null constraint when inserted into the Virasoro constraints is to add in the twisted sectors with fractional winding.  This shift orbifold is nothing but the discrete remnant of the gauge motion~\eqref{Horbit} left after gauge fixing $t$ and $y$, as discussed above.

\subsection{General embeddings $\cH\hookrightarrow\cG$}
\label{sec:generalembed}

More generally, the shift orbifold acts on $\sltwo\times\sutwo$ as an asymmetric shift orbifold~\rcite{Martinec:2018nco}.%
\footnote{Asymmetric orbifolds are discussed for instance in~\rcite{Narain:1986qm,Narain:1990mw}.}
Parametrizing the group manifold via Euler angles%
\footnote{We find it convenient to work with $SU(1,1)$ rather than $\sltwor$, as that diagonalizes the symmetries being gauged.}
\be
\label{eulerangs}
g_\sl = e^{\frac i2(\tau-\sigma)\sigma_3} \, e^{\rho\sigma_1}\, e^{\frac i2(\tau+\sigma)\sigma_3}
~~,~~~~
g_\su = e^{\frac i2(\phi-\psi)\sigma_3} \, e^{i(\frac\pi2-\theta)\sigma_1}\, e^{-\frac i2(\phi+\psi)\sigma_3}  ~,
\ee
the orbifold identifies the group manifold under the shift~\rcite{Jejjala:2005yu,Chakrabarty:2015foa} 
\be
\label{orbact}
\delta(\sigma,\phi,\psi) = \frac{2\pi}{\sfk} \big( 1,\sfn,\sfm \big) ~.
\ee
which in the bosonized representation~\eqref{bosonize}, \eqref{vertexops} amounts to
\be
\label{asymshiftorb}
\delta\big(\cY_\sl-\bar \cY_\sl\big)  = \frac{2\pi}{\sfk}  
~~,~~~~
\delta\cY_\su  = -(\sfn+\sfm)\frac{2\pi}{\sfk} 
~~,~~~~
\delta \bar\cY_\su = (\sfn-\sfm)\frac{2\pi}{\sfk} ~.
\ee

The null constraints are now
\begin{align}
\begin{split}
\label{genlargeRnull}
\sfk(\vareps+n_y) &= 2m+n_5 w + 2M - (\sfm+\sfn)(2m'+n_5w'+M') -\Big[\frac{(\sfm+\sfn)^2-1}{2\sfk}\Big]n_5w_y
\\[.2cm]
\sfk(\vareps-n_y) &= 2\bar m+n_5 w + 2\bar M - (\sfm-\sfn)(2\bar m'+n_5\bar w'+\bar M') -\Big[\frac{(\sfm-\sfn)^2-1}{2\sfk}\Big]n_5w_y ~,
\end{split}
\end{align}
while the Virasoro constraints are as before, equation~\eqref{largeRvir}.  The axial null constraint at large $R_y$ imposes
\be
-\sfk n_y = 
m-\bar m+M-\bar M - (\sfm+\sfn)\Big(m'+\frac{n_5}{2}w'+M' \Big) + (\sfm-\sfn)\Big( \bar m'+\frac{n_5}{2}\bar w'+\bar M' \Big) - \frac{\sfm\sfn n_5}{\sfk}\,w_y  ~.
%
\ee
In the sector $w_y=0$, this constraint imposes that some linear combination of left/right $\sltwo$ and $\sutwo$ $J^3$ charges is a multiple of $\sfk$ (recall that $\sfm\sfn n_5/\sfk\in\bZ$).  This is the untwisted sector of the orbifold; when $\sfn\ne 0$ the orbifold action is left/right asymmetric on the worldsheet.  Note that one can regard this untwisted sector constraint as imposing on the momenta along $\sigma,\phi,\psi$ the condition that $P_\sigma+\sfn P_\phi + \sfm P_\psi$ be a multiple of $\sfk$, which is precisely what one would deduce from the identification~\eqref{orbact}.  The twisted sectors of this asymmetric orbifold are again labelled by $w_y$, and amount to the fractional spectral flow
\be
w\to w + w_y/\sfk
~~,~~~~
w'\to w' + (\sfm+\sfn)w_y/\sfk
~~,~~~~
\bar w' \to \bar w' + (\sfm-\sfn)w_y/\sfk  ~,
\ee
which amounts to adding strings that close only under the identifications~\eqref{orbact}.

The state in the dual CFT has quantum numbers 
\begin{align}
\begin{split}
\label{qnums}
h = \frac{N}{4}\bigg( 1+\frac{(\sfm+\sfn)^2-1}{\sfk^2} \bigg)  
~~&,~~~~
J_\cR = \frac{N}{2}\,\frac{\sfm+\sfn}{\sfk}
\\[.2cm]
\bar h = \frac{N}{4}\bigg( 1+\frac{(\sfm-\sfn)^2-1}{\sfk^2} \bigg)  
~~&,~~~~
\bar J_\cR = \frac{N}{2}\,\frac{\sfm-\sfn}{\sfk}
\end{split}
\end{align}
where $J_\cR$ is the $J^3$ component of the $\sutwo$ $\cR$-symmetry (similarly for $\bar J_\cR$).
The state is 1/4-BPS (excited on the left) when $\sfm-\sfn=\pm1$, and 1/2-BPS when $\sfn=0$, $\sfm=\pm1$.  We will mostly be interested in these cases, for which the states are in the R-R sector of the spacetime CFT~\rcite{Jejjala:2005yu,Giusto:2012yz,Chakrabarty:2015foa}.  More generally, the states are in the R-R sector for $\sfm+\sfn$ odd, and in the NS-NS sector for $\sfm+\sfn$ even; the latter are all non-BPS.

\subsection{States in the dual CFT} 
\label{sec:holomap}

The BPS states of the NS5-F1 system are simple to describe in terms of the T-dual NS5-P system.  There, they are simply BPS waves on the fivebranes.  In a sector where the fivebranes are twisted into a single fivebrane wrapping $n_5$ times around $\bS^1_y$, momenta are fractionated by a factor of $n_5$; one can then have any number $n_k^I$ of modes with momenta $k/N$ in any of 8 bosonic and 8 fermionic polarizations $I$, subject to the overall constraint 
\be
\label{sumrule}
\sum_{k,I} k \, n_k^I = N  ~.
\ee
We can thus label the 1/2-BPS spectrum via
\be
\label{halfbpsstate}
\big| \{n_k^I\}\big\rangle ~,
\ee
and this labeling passes through the T-duality to the NS5-F1 frame where the labels refer to a collection of (generically fractional, unless $k$ is a multiple of $n_5$) winding strings carried as fivebrane excitations.  The bosonic excitations split into the four scalars $X^{\alpha\dot\alpha}$ that describe the gyration of the fivebrane in its transverse space, and four more comprising a gauge multiplet on the fivebrane.%
\footnote{For type IIB in the NS5-F1 frame, the polarization labels refer to the gauge multiplet $A^{AB}$ on the T-dual fivebrane consisting of a scalar and a self-dual antisymmetric tensor.  The scalar $A^{[AB]}$ is typically referred to in the literature as the ``00'' mode.}
The $\bZ_\sfk$ orbifold geometries are dual to the states
\be
\label{Zkstates}
\Big| \big\{n_\sfk^{++}\tight=N/\sfk, {\it others} \tight=0 \big\} \Big\rangle ~.
\ee

The 1/2-BPS supergravity solutions of the NS5-F1 system can be put in a standard form~\cite{Lunin:2001fv,Kanitscheider:2007wq} (restricting for simplicity to solutions with pure NS-NS fluxes; for the general solution, see \eg\ Appendix B of~\rcite{Martinec:2022okx})
\begin{align}
\begin{aligned}
\label{LMgeom}
ds^2 &\;=\; \K  ^{-1}\bigl[ -(d\tau+\A)^2 + (d\sigma + \B)^2 \bigr] + \H \, d\xx\!\cdot\! d\xx + ds^2_\cM \,,
\\[.2cm]
B &\;=\; 
\K  ^{-1}\bigl(d\tau + \A\bigr)\wedge\bigl(d\sigma+\B\bigr)  + \C_{ij}\, dx^i\wedge dx^j \,,
\\[.2cm]
e^{2\Phi} &\;=\; 
{\gstrsq}\,\frac{\H}{\K}  \,, \qquad~~~ d\C =  *_{\sst\perp} d\H \;, 
\qquad~~ d\B = *_{\sst\perp} d\A \,,
\end{aligned}
\end{align}
where $\xx$ are Cartesian coordinates on the transverse space to the fivebranes, related to $\sltwo\times\sutwo$ Euler angles via
\be
\label{bipolar}
x^1+ix^2 \equiv x^{++} = \cosh\rho\,\sin\theta \, e^{i\phi}
~~,~~~~
x^3+ix^4 \equiv x^{-+} = \sinh\rho\,\cos\theta\, e^{i\psi}  ~.
\ee

The harmonic forms and functions appearing in this solution can be written in terms of a Green's function representation,
which in the $AdS_3$ decoupling limit takes the form
\begin{align}
\begin{aligned}
\label{greensfn}
\H  \,=\, \frac{1}{2\pi}\sum_{m=1}^\nfive\int\limits_0^{2\pi} \frac{d\tilde{v}}{|\xx-\F_\flabel(\tilde{v})|^2} ~, & \qquad~
\K   \,=\,  
\frac{1}{2\pi}\sum_{m=1}^\nfive\int\limits_0^{2\pi} \frac{d\tilde{v} \, \dot\F_\flabel \tight\cdot \dot\F_\flabel}{|\xx-\F_\flabel(\tilde{v})|^2} \;,\\[.2cm]
\A \,=\, \A_i dx^i \,, \qquad
\A_i \,=\, & \frac{1}{2\pi}\sum_{m=1}^\nfive\int\limits_0^{2\pi} \frac{d\tilde{v} \,\dot \F^\flabel_i(\tilde{v})}{|\xx-\F_\flabel(\tilde{v})|^2}\;,
\end{aligned}
\end{align}
involving source profile functions $\F^i_\flabel(\tilde{v})$, $m=1,2,\dots,\nfive$, that describe the locations of the fivebranes in their transverse space (overdots denote derivatives with respect to $\tilde{v}$).
We choose twisted boundary conditions for the source profile functions,
\be
\label{twistedbc}
\F^i_\flabel(\tilde{v}+2\pi)=\F^i_{\sst (m+1)}(\tilde{v}) 
\ee
that bind all the fivebranes together (provided $\sfk$ and $n_5$ are relatively prime) and introduce the fractional moding described above.  We can then bundle all the fivebrane profile functions together into a single profile $\F^I(\tilde{v})$ that extends over the range $[0,2\pi\nfive)$.

\begin{figure}[ht]
\centering
\includegraphics[width=0.4\textwidth]{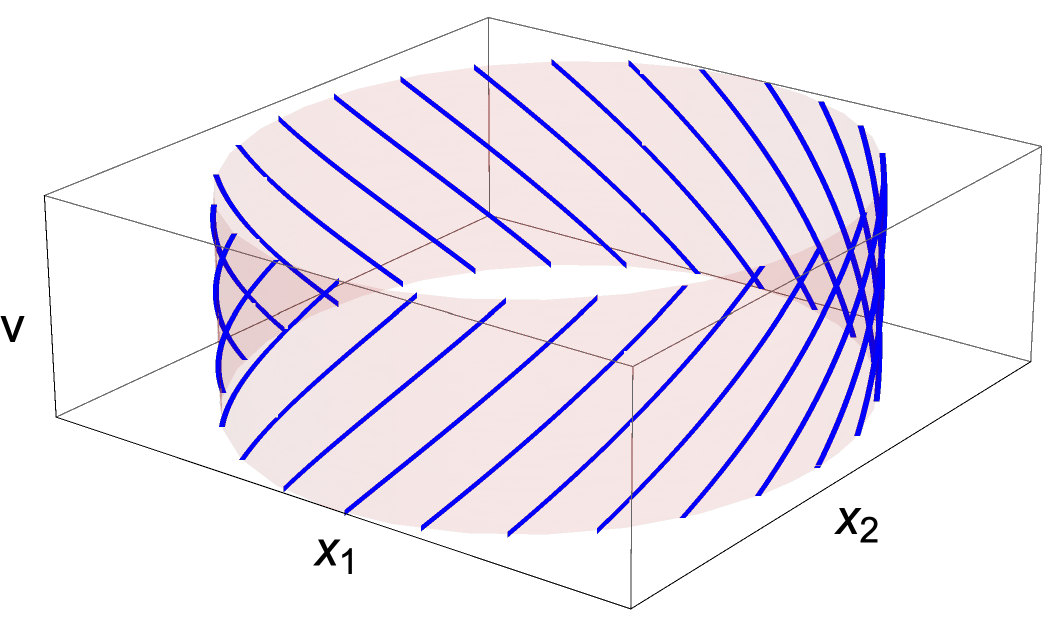}
\caption{\it Circular supertube source profile, in which only a single mode is excited (in this case, $\sfk = 3$ and $n_5 = 25$), so that the fivebranes spiral around a torus in $(y, x^1, x^2)$ shaded in pink. }
\label{fig:STspiral}
\end{figure}

The key point here is that the Fourier amplitudes $a_k^I$ of the source profile functions 
\be
\F^I(\tilde v) = \sum_k \frac{\sfa_k^I}{\sqrt{k}}\, e^{ik\tilde v/n_5}
\ee
are coherent state parameters whose expectation values are the mode numbers $n_k^I$.  We thus have a direct map between the labels of 1/2-BPS states and the bulk geometries they correspond to, at the fully non-linear level.  In particular, for the orbifold geometries $(AdS_3\times\bS^3)/\bZ_\sfk$, we know exactly what the fivebranes are doing~-- the source profile function has only a single mode excited
\be
\label{Zkprofile}
\F^{++}(\tilde v) = \frac{\sfa^{++}_\sfk}{\sqrt{k}} \, e^{i\sfk \tilde v/n_5} 
\ee
with $\vev{\sfa^{++}_{-\sfk}\,\sfa^{++}_\sfk} = N/\sfk$, and describes fivebranes spiraling around in the torus parametrized by the T-dual to the $AdS_3$ azimuthal coordinate $\sigma$ and the $\bS^3$ Euler angle $\phi$, and sitting at $\rho=0$~\rcite{Martinec:2017ztd}.  The spiral runs along the $(\sfk,n_5)$ cycle of this torus; see figure~\ref{fig:STspiral}.

%
%

While supergravity only sees the smeared average of the fivebrane/string source, worldsheet string theory is sensitive to the underlying fivebrane source distribution.  The spiraling source distribution of the orbifold states is seen by D-brane probes that end on the fivebranes~\rcite{Martinec:2019wzw}.  There is a $\bZ_{n_5}$ structure to the D-branes of the $\sutwo$ WZW model at level $n_5$ (see for instance~\rcite{Maldacena:2001ky}), and a $\bZ_\sfk$ symmetry of fractional brane states on the orbifold~\rcite{Diaconescu:1997br}; together these should realize, directly in the $R_y\to\infty$ limit, the spiraling structure seen in the somewhat more elaborate construction of boundary states in the gauged WZW model at general $R_y$ given in~\rcite{Martinec:2019wzw}.

The CFT dual is often described in the language of the symmetric product orbifold, which pertains to a weak-coupling region of the moduli space.  In the symmetric product, 1/2-BPS states are associated to conjugacy classes of the symmetric group, which are labeled by the same data~\eqref{sumrule}, \eqref{halfbpsstate}, and describe collections of copies (cycles) of the block $\bT^4$ CFT that are sewn together by a cyclically twisted boundary condition analogous to~\eqref{twistedbc}.  Each cycle has a collection of 1/2-BPS ground states labeled by the same data as the polarization labels $I$ carried by the bulk fivebrane excitations.

The BPS states are preserved under the marginal deformation to the strongly-coupled regime of the CFT where the bulk dual has a supergravity approximation.  The analysis of~\rcite{Martinec:2020gkv,Martinec:2022okx} shows that much of the symmetric product structure survives this deformation.  In particular, the effect of 1/2-BPS string vertex operators is to deform perturbatively the string winding condensate carried by the fivebranes, and at the same time the geometry that the condensate is sourcing.%
\footnote{The aspects of the vertex operator that are responsible for these two effects are related by FZZ duality~\rcite{Giveon:2016dxe,Martinec:2020gkv}.}
For instance, a 1/2-BPS graviton vertex operator $\cV^{\alpha\dot\alpha}_{j',w_y}$ sews together a number of background strings into a longer string, while changing its polarization state:
\be
\label{Vtransition}
\big(|\tight++\rangle_\sfk\big)^{2j'+1}~\longrightarrow~ |\alpha\dot\alpha\rangle_{(2j'+1)\sfk+w_y n_5}
\ee
where $|I\rangle_k$ denotes a cycle of length $k$ in polarization state $I$
(see~\rcite{Martinec:2020gkv,Martinec:2022okx} for details).
Exponentiating the vertex operators thus coherently changes the winding condensate carried by the fivebranes as specified by the profile functions $\F^I(\tilde v)$.

\subsection{Deformation back to linear dilaton asymptotics} 
\label{sec:lindil}

We have shown that the $R_y\to\infty$ endpoint of the marginal line of null gauged WZW models~\eqref{GmodH} can be described as a shift orbifold of the $\sltwo\times\sutwo$ theory.  The full marginal line should therefore have a description as a marginal deformation of the orbifold.  The obvious candidate is a deformation by the bilinear of the L/R currents that generate the shift.  To end this section, we show that indeed the null gauged WZW model is the marginal line whose linearized deformation away from the orbifold $\cK/\bZ_\sfk$ is by the operator $\cJ\bar\cJ$.%
\footnote{Current-current marginal deformations were explored in~\rcite{Hassan:1992gi,Giveon:1993ph,Gershon:1993wu}, and the review~\rcite{Giveon:1994fu}.}

The null-gauged WZW model yields the bulk NS5-F1 circular supertube supergravity solution at finite $R_y$~\rcite{Martinec:2017ztd,Bufalini:2021ndn,Martinec:2022okx} 
\begin{align}
\label{NS5F1}
ds^2 &= \Bigl( -du\:\! dv + ds_{\scriptscriptstyle\mathbf T^4}^2  \Bigr)
+ {n_5}\Bigl[ d\rho^2+ d\theta^2+\frac1\Sigma \Bigl( {\cosh}^2\!\rho\,\sin^2\!\theta \,d\phi^2 + {\sinh}^2\!\rho\,\cos^2\!\theta \,d\psi^2 \Bigr)\Bigr] 
\nonumber\\[.1cm]
& \hskip .6cm 
+ \frac{2\alphab}{\Sigma} \Bigl( \cos^2\!\theta \, dy\,d\psi - {\sin^2\!\theta \, dt\,d\phi}  \Bigr)
+\frac{\alphab^2}{\nfive\Sigma} \Bigl[ \nfive \sin^2\!\theta \, d\phi^2 +  \nfive\cos^2\!\theta \, d\psi^2  
+ du\, dv \Bigr],
\nonumber\\[8pt]
B  &= \frac{ \cos^2\!\theta (\alphab^2+\nfive\cosh^2\!\rho)}{\Sigma}  { d\phi\wedge d\psi + \frac{\alphab^2}{n_5\Sigma} \, dt\wedge dy } \nonumber\\
& \hskip .6cm 
{}-\frac{\alphab  \cos^2\!\theta}{\Sigma}  dt\wedge d\psi
+ \frac{\alphab  \sin^2\!\theta}{\Sigma}  dy\wedge d\phi~, 
\qquad\qquad\quad u=t+y \;, ~~v=t-y \;,
\nn\\[8pt]
e^{-2\Phi} & = \frac{n_1\Sigma}{\sfk^2\Ry^2\,V_4} ~,\qquad~~~  
\Sigma = \frac{\alphab^2}{\nfive} + \sinh^2\rho + \cos^2\theta ~, \qquad\quad \alphab\equiv \sfk\Ry 
\end{align} 
in the gauge $\tau=\sigma=0$;
it has a $\bZ_\sfk$ orbifold singularity located along the circle $\rho=0,\theta=\pi/2$ parametrized by $\phi$.
The generalization to the geometries resulting from the generic null vectors~\eqref{nullvecs} may be found for instance in~\rcite{Martinec:2017ztd,Martinec:2018nco,Bufalini:2021ndn}, and the CFT dual is described in~\rcite{Giusto:2012yz,Chakrabarty:2015foa}.

The limit $R_y\to\infty$ of the background~\eqref{NS5F1} is the orbifold~\eqref{orbact}, as we saw by choosing to gauge fix $t,y$ instead of $\tau,\sigma$, and noting that a residual discrete gauge symmetry implements the orbifold identification on $\cK$ as discussed in~\rcite{Martinec:2018nco}.  Backing away from the large $R_y$ limit, null gauging modifies the group sigma model on $\cG$ by a term
\be
\label{JJbardef}
\frac{\cJ\bar\cJ}{\Sigma}
\ee
after integrating out the gauge field.  In the gauge $t=y=0$, this is just a current-current deformation of $\cK/\bZ_\sfk$.  Indeed, in this gauge the currents $\cJ,\bar\cJ$ reduce to their components along $\cK=\sltwo\times\sutwo$, and $\Sigma^{-1}\sim n_5/(\sfk R_y)^2$ is the parameter of the infinitesimal current-current deformation.%
\footnote{This relation between the null-gauged model and current-current deformations was also observed in~\rcite{Bufalini:2021ndn}.}
This result is a special case of the analysis of~\rcite{Forste:2003km} relating gauged WZW models of the form $(\cK\times\cH)/\cH$ to current-current deformations of orbifolds of $\cK$.

A special case of the above orbifolds was considered in~\rcite{Martinec:2001cf,Martinec:2002xq}, where a restriction was made to orbifold orders $\sfk$ that are divisors of $n_5$ under the thinking that other choices would be anomalous.  We see here that this restriction can be lifted and that the orbifold is always consistent.


\section{Generalization: Moving defects} 
\label{sec:conjugation}

The static conical defect geometries above arise when we diagonalize the action of $\bZ_\sfk$ in $\sltwo$ so that it is an elliptic transformation that fixes the timelike geodesic at the center of $AdS_3$, \ie\ $\rho=0$ in the $AdS_3$ global coordinates~\eqref{eulerangs}.
  
Elements of $\sltwo$ lie in one of three conjugacy classes.  Realizing group elements $g\in\sltwo$ as $2\times2$ matrices, the conjugacy classes are characterized by the matrix trace
\begin{align}
\begin{split}
{\rm Elliptic}~&:~~ \big| \tr(g)\big| < 2 ~~,
\\
{\rm Parabolic}~&:~~ \big| \tr(g)\big| = 2 ~~,
\\
{\rm Hyperbolic}~&:~~ \big| \tr(g)\big| > 2 ~~.
\end{split}
\end{align}
But group theoretically, we can conjugate a discrete group identification $g\sim \gamma_R^{-1}\,g\,\gamma_L^{~}$ by any elements $h_R,h_L$ on the left and right to find an equivalent $\bZ_\sfk$ identification  
\be
\label{conjugated}
g ~\sim ~ (h_R^{~}\gamma_R^{-1} h_R^{-1})\,g\,(h_L^{-1} \gamma_L^{~}h_L^{~}) ~.
\ee
The conjugated transformations needless to say lie in the same conjugacy class as the originals, of course.

For instance, we can take $h_L^{~},h_R^{~}$ to be boost transformations in $\sltwo$ (\ie\ hyperbolic group elements).
These do not leave fixed the time translation Killing vector $\partial_\tau$, and so unlike the original orbifold action which was time-translation invariant, the new orbifold identification will be time-dependent.  One finds in this way a conical defect moving along a massive particle geodesic, which oscillates and rotates around the center of $AdS_3$.  Such a boosted defect (having $\gamma_L^{~}=\gamma_R^{~}$, and thus moving purely radially) is depicted in figure~\ref{fig:DefectBoost}; generalizing to arbitrary $\gamma_L^{~}\ne\gamma_R^{~}$ results in an oscillating, rotating defect of the sort depicted in figure~\ref{fig:DefectRot}.

Note that these states are distinct from the ``superstrata'' or ``microstrata'' considered in~\rcite{Bena:2015bea,Bena:2017xbt,Ganchev:2021pgs}.  In those constructions one is making a coherent excitation by strings that individually carry $AdS_3$ momentum excitations $\cL_{-1},\bar \cL_{-1}$~\rcite{Martinec:2022okx}, while here we are making a coherent excitation acting by $\exp[z \cL_{-1} + \bar z \bar \cL_{-1}]$ on the whole state.

The inclusion of $\sutwo$ transformations $(\gamma_L',\gamma_R')$ in the conjugation of the orbifold identification has the additional effect of rotating the orientation of the source ring along $\bS^3$.  The initial defect lies along the ring $\rho=0,\theta=\pi/2$, and is extended along $\tau$ and $\phi$.  The conjugation rotates the ring onto some other great circle in $\bS^3$.  For simplicity. we will not consider this possibility further.

The static geodesic in the center of $AdS_3$ is given by $g=e^{2\pi i\nu\xi\sigma_3}$.
A boosted defect travels along the $\sltwo$ geodesic
\be
\label{geoboost}
g(\xi) = e^{\etaR \sigma_1/2} \, e^{2\pi i\nu\xi\sigma_3} \, e^{\etaL \sigma_1/2} ~,
\ee
having classical conserved charges
\begin{align}
\label{charges}
\cE &= \frac{N}{16\pi^2}\tr\bigl[\partial_\xi g \partial_\xi g^{-1}\bigr] = \frac{N}2 \nu^2
\nn\\[.2cm]
J^3_\sl &= -\frac{iN}{4\pi} \tr\bigl[g^{-1}(\partial_\xi g)\sigma_3\bigr] = N\nu\cosh(\etaL)
\\[.2cm]
\bar J^3_\sl &= -\frac{iN}{4\pi}  \tr\bigl[(\partial_\xi g)g^{-1}\sigma_3\bigr] = N\nu\cosh(\etaR)  ~~.
\nn
\end{align}
The $AdS_3$ angular momentum is nonzero when $\etaL\ne \etaR$.
The parameter $\nu$ determines the deficit angle $\alpha$ and rest mass $\ell M$ of the defect as
\be
\alpha = 2\pi\nu
~~,~~~~
\ell M = \frac N2\big( \nu^2 -1\big)
\ee
where $\ell$ is the $AdS_3$ curvature radius and
\be
N = \frac{\ell}{4G} = \frac{ c_{ST}^{~}}{6} = n_1n_5 ~,
\ee
with $c_{ST}^{~}$ the central charge of the dual spacetime CFT and $G$ the 3d Newton constant.
In the worldsheet theory, the boosted defect is simply the quotient of $\sltwo\times\sutwo$ by a conjugate embedding of $\bZ_\sfk$.

Multiple conical defects traveling along geodesics in $AdS_3\times \bS^3$ can be described classically by more general orbifolds, generated by a collection of boosted $\bZ_{\sfk_i}$ identifications, where $i$ labels the defect.  
The set of generating transformations $\{\gamma_{L,i},\gamma_{R,i}\}\in\cK$ is individually an elliptic transformation making a conical defect.  Each defect can be locally BPS, but the supersymmetries are misaligned by the relative boosts and so all supersymmetries are broken.  The holonomy around subsets of defects is given by the products of the left and right group elements associated to the members of the subset.  The product is always hyperbolic~-- the holonomy of the BTZ black hole.%
\footnote{For defects of small deficit angles, which are not global orbifolds of $\sltwo\times\sutwo$, the holonomy around aggregates of conical defects can remain elliptic; but the deficit angle around global orbifold defects is at least $\pi$, and we will see that the holonomy around even a pair of such defects is hyperbolic.}

\subsection{Non-orbiting defects and non-rotating black holes}
\label{sec:radialboost}

\begin{figure}[ht]
\centering
  \begin{subfigure}[b]{0.3\textwidth}
  \hskip 0cm
    \includegraphics[width=\textwidth]{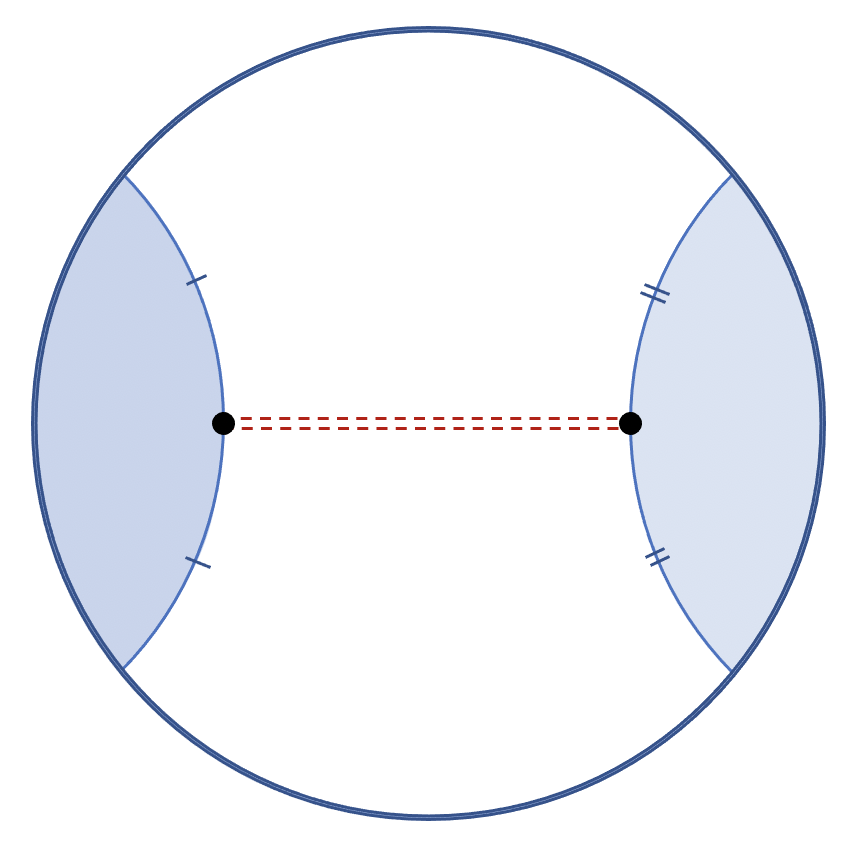}
    \caption{ }
    \label{fig:Z2cutout}
  \end{subfigure}
\qquad\qquad
  \begin{subfigure}[b]{0.3\textwidth}
      \hskip 0cm
    \includegraphics[width=\textwidth]{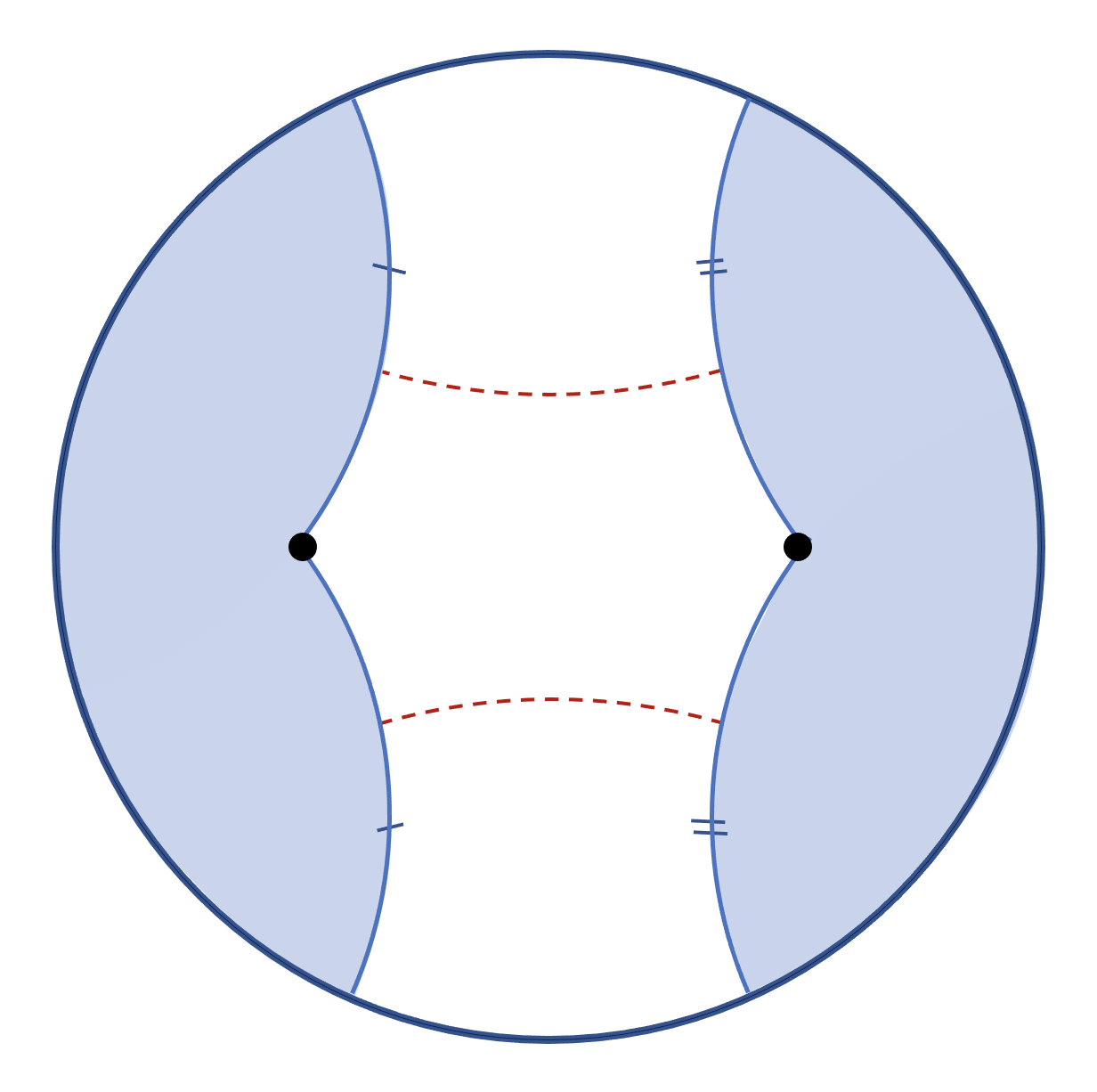}
    \caption{ }
    \label{fig:Z3cutout}
  \end{subfigure}
\caption{\it 
a) The spatial geometry of two $\bZ_2$ defects at the moment of time reflection symmetry, when both defects are at rest.  The shaded regions are excised, and the circular arcs on either side of a given defect are identified.  The red dashed line is the apparent horizon, which is degenerate in this example (\ie\ the defects are at the apparent horizon at the surface of time reflection symmetry).\\
$~~~~$
b) The spatial geometry of two $\bZ_3$ defects at the moment of time reflection symmetry.  The red dashed line is the apparent horizon; the defects lie behind it as seen from afar.
}
\label{fig:cutout}
\end{figure}

Consider for instance two $\bZ_\sfk$ defects, boosted by equal and opposite amounts $\pm\eta$ on both left and right.  The spatial hypersurface at $\tau=0$ is the Poincar\'e disk $\bH_2$, with the defects momentarily at rest at some radius $\rho=\eta$.  The disk has two wedges cut out with defect angle $\alpha$; the sides of the wedges are geodesics in $\bH_2$ passing from the defects to the boundary of the disk, see figure~\ref{fig:cutout}.  As time evolves, the two defects fall toward each other and collide.  Beyond this collision point is a region of timelike identification in $\sltwo$.  This region of closed timelike curves is usually excised from spacetime as being unphysical.

The two defects have identifications under 
\be
\label{combined}
\gsl\sim \gamma_{R,i}^{-1} \,\gsl\, \gamma_{L,i}^{~}
~~,~~~~
\gsu\sim (\gamma'_{R,i})^{-1} \,\gsu\, \gamma'_{L,i}
~~,~~~~ i=1,2~,
\ee 
where we take (again for simplicity) $\gamma'_{R,i}=\gamma'_{L,i}=e^{i\pi\sigma_3/\sfk}$.
For both defects having deficit angle $\alpha$ one has%
\footnote{More generally, one could displace the defect in the direction with azimuthal angle $\phi$, in which case one should rotate the group elements to $\gamma_{L,R}^{~} \to e^{-i\phi\sigma_3/2} \gamma_{L,R}^{~} e^{i\phi\sigma_3/2}$.}
\begin{align}
\begin{split}
\label{2Z2}
\gamma_{L,1}^{~} = \gamma_{R,1}^{~} = e^{-\eta\sigma_1/2} \, e^{+i\alpha\sigma_3/2} \,e^{+\eta\sigma_1/2}
~~&,~~~~
\gamma_{L,2}^{~} = \gamma_{R,2}^{~} = e^{+\eta\sigma_1/2} \, e^{+i\alpha\sigma_3/2} \,e^{-\eta\sigma_1/2}
\end{split}
\end{align}
The product of these two transformations specifies the holonomy of the connection around the pair of defects.  The conjugacy class is given by
\be
\label{holonomy}
 \tr( \gamma_{L,1}^{~} \,\gamma_{L,2} ) =  \tr( \gamma_{R,2}^{~} \,\gamma_{R,1} ) = 2(\cos\alpha\,\cosh^2\!\eta-\sinh^2\!\eta)
\ee
For example, two $\bZ_2$ defects ($\alpha=\pi$) on top of one another ($\eta=0$) make an extremal black hole (the parabolic conjugacy class $|Tr(g)|=2$); for any finite separation $\eta>0$, the holonomy around a pair of $\bZ_2$ defects is in the hyperbolic conjugacy class $|Tr(g)|>2$ that characterizes a BTZ black hole.%
\footnote{Exceptionally, one can regard the geometry of a pair of $\bZ_2$ defects as the $\bZ_2$ orbifold of the vacuum BTZ geometry itself, by the reflection that exchanges the two exterior regions; this explains why the defects are always on the horizon at the moment of time reflection symmetry.}  
Smaller defects (which cannot arise as global orbifolds) can collide to make either composite conical defects of larger deficit angle, or black holes; defects of larger deficit angle than $\pi$ (and thus all of the $\bZ_\sfk$ defects for $\sfk>2$) always make black holes.

\begin{figure}[ht]
\centering
  \begin{subfigure}[b]{0.3\textwidth}
  \hskip 0cm
    \includegraphics[width=\textwidth]{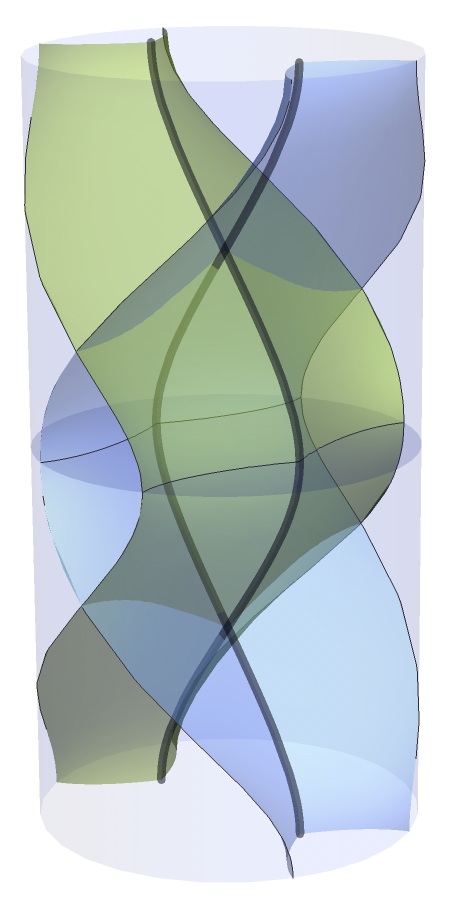}
    \caption{ }
    \label{fig:LorZ3-3D}
  \end{subfigure}
\qquad\qquad
  \begin{subfigure}[b]{0.31\textwidth}
      \hskip 0cm
    \includegraphics[width=\textwidth]{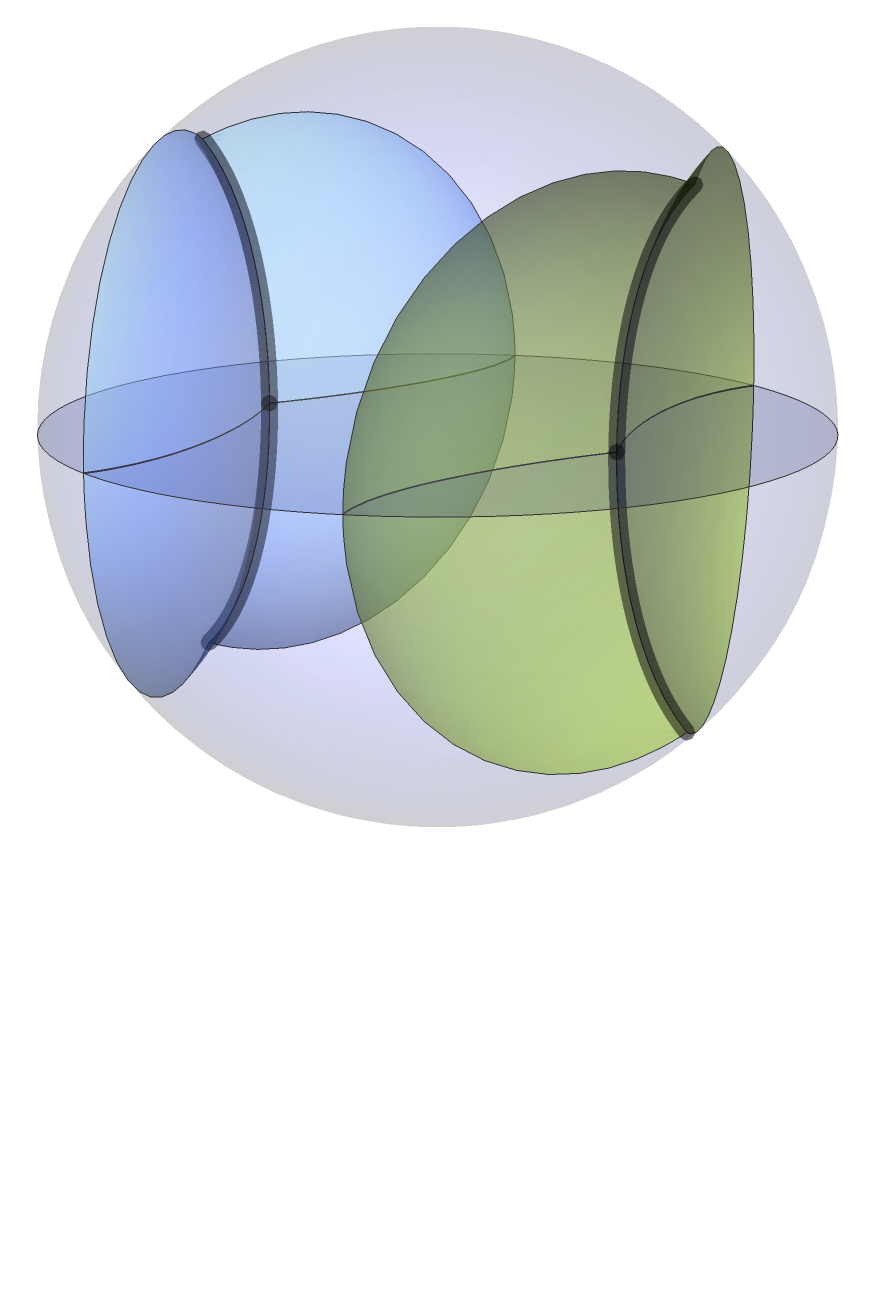}
    \caption{ }
    \label{fig:EucZ3-3D}
  \end{subfigure}
\caption{\it 
a) The collision of two $\bZ_3$ defects, obtained by identifying global $AdS_3$ by separate $2\pi/3$ rotations about oppositely boosted geodesics (the thick black lines), makes a BTZ black hole.  The surfaces being identified on either side of the geodesic are shaded blue for one defect and green for the other for ease of viewing.  The past and future BTZ singularities are the locus where the sheets being identified collide; the region between the sheets is the physical BTZ geometry (the intersection of the front side of the blue sheet and the back side of the green sheet).  Note that due to the identifications, there is only one connected asymptotic region.  The future (past) horizons are the past (future) light cones of the points where the singularity meets the conformal boundary.  The defects are always well inside the horizon.\\
$~~~~$
b) Euclidean continuation of two $\bZ_3$ defects.  
The geometry on the surface of time reflection symmetry agrees with the Lorentzian section and is depicted in figure~\ref{fig:Z3cutout}.
}
\label{fig:DefectCollisions}
\end{figure}

Indeed, the BTZ solution can also be described as a $\bZ$ orbifold of $\sltwo$ generated by the hyperbolic transformation
\be
\gsl~\sim~ e^{\pi(r_+-r_-)\sigma_1} \, \gsl \, e^{\pi(r_++r_-)\sigma_1}
\ee
where $r_\pm$ are the inner and outer horizon radii relative to the $AdS$ scale $\ell$, in terms of which the black hole mass and angular momenta are given by  
\be
\label{MJrpm}
\ell M = \frac{\ell (r_+^2+r_-^2)}{8G} = \frac N2 \big( {r_+^2 + r_-^2} \big)
~~,~~~~
J = \frac{\ell \, r_+ r_-  }{4G} = N\, r_+ r_- ~.
\ee 
For the non-spinning black hole that forms from a pair of $\bZ_2$ conical defects via~\eqref{2Z2}, one has from equation~\eqref{holonomy} $r_-=0$ and $r_+=2\eta/\pi$.  Essentially, the defects' separation at the moment of time symmetry becomes kinetic energy when they collide, so the greater the separation on this hypersurface, the greater the mass of the final state black hole.  The collision of a pair of $\bZ_2$ defects is depicted in figure~\ref{fig:Z2-BTZ}, and that of two $\bZ_3$ defects in figure~\ref{fig:LorZ3-3D}.

These backgrounds seem quite bizarre in Lorentz signature~-- they consist of a past black hole singularity that spontaneously spits out a pair of conical defects, which are then re-absorbed into a future singularity.  Here we have perhaps a black hole analogue of all the air in a room rushing to one corner, in the sense that the random chaotic distribution of degrees of freedom that characterizes black hole microstates somehow spontaneously organizes into a tiny coherent corner of phase space (the pair of defects), and then just as quickly devolves into a random chaotic mess again.  This process is not visible to the external observer, for whom the entire sequence is hidden behind the event horizon.%
\footnote{There may be a horizon distortion away from spherical symmetry, since rotational symmetry in $\sigma$ would seem to broken to the $\bZ_2$ rotation that permutes the defects.  This would be seen as a set of quasinormal modes decaying along the future horizon, having emerged from the past horizon in similar fashion.}
The singularity can be characterized as the locus where the transformation~\eqref{combined} becomes null.%
\footnote{Beyond this surface is a region of timelike identifications that is not usually considered part of spacetime, but is part of the Lorentzian orbifold geometry.  For this reason, we will soon pass to Euclidean $AdS_3$ where the group action is always sensible.}
The past (future) light cone of the intersection of the future (past) singularity and the conformal boundary is the future (past) horizon.  The singularity locus on the conformal boundary is a set of fixed points of the identification.  For the pair of $\bZ_2$ defects~\eqref{2Z2}, the intersection of the future singularity with the conformal boundary lies at $\tau=\pi/2$, $\sigma=\pm\pi/2$ (which are the same point under the orbifold identification), and for the past singularity at $\tau=-\pi/2$, $\sigma=\pm\pi/2$. 

The work of~\rcite{Steif:1995pq,Birmingham:1999yt,Brill:2007zq,Lindgren:2015fum} shows that the defects lie inside an apparent horizon whenever there is a closed geodesic that surrounds them on the surface of time symmetry.  The geometry of this surface of time symmetry is the Poincar\'e disk $\bH_2$; geodesics are circles that intersect the boundary of this disk orthogonally.  The straight line between the two defects in figure~\ref{fig:Z2cutout} is such a geodesic; thus we see that a path that travels along the upper side of this geodesic from the first defect to the second one, follows the identification and returns slight below the geodesic to the first defect, is a closed path that is a geodesic in the limit that the displacement above and below shrink to zero.  Thus the pair of $\bZ_2$ defects lies along a degenerate apparent horizon.  In figure~\ref{fig:Z3cutout}, the red dashed line is a closed geodesic due to the identifications; thus the defects lie within the apparent horizon.

\begin{figure}[ht]
\centering
\includegraphics[width=0.4\textwidth]{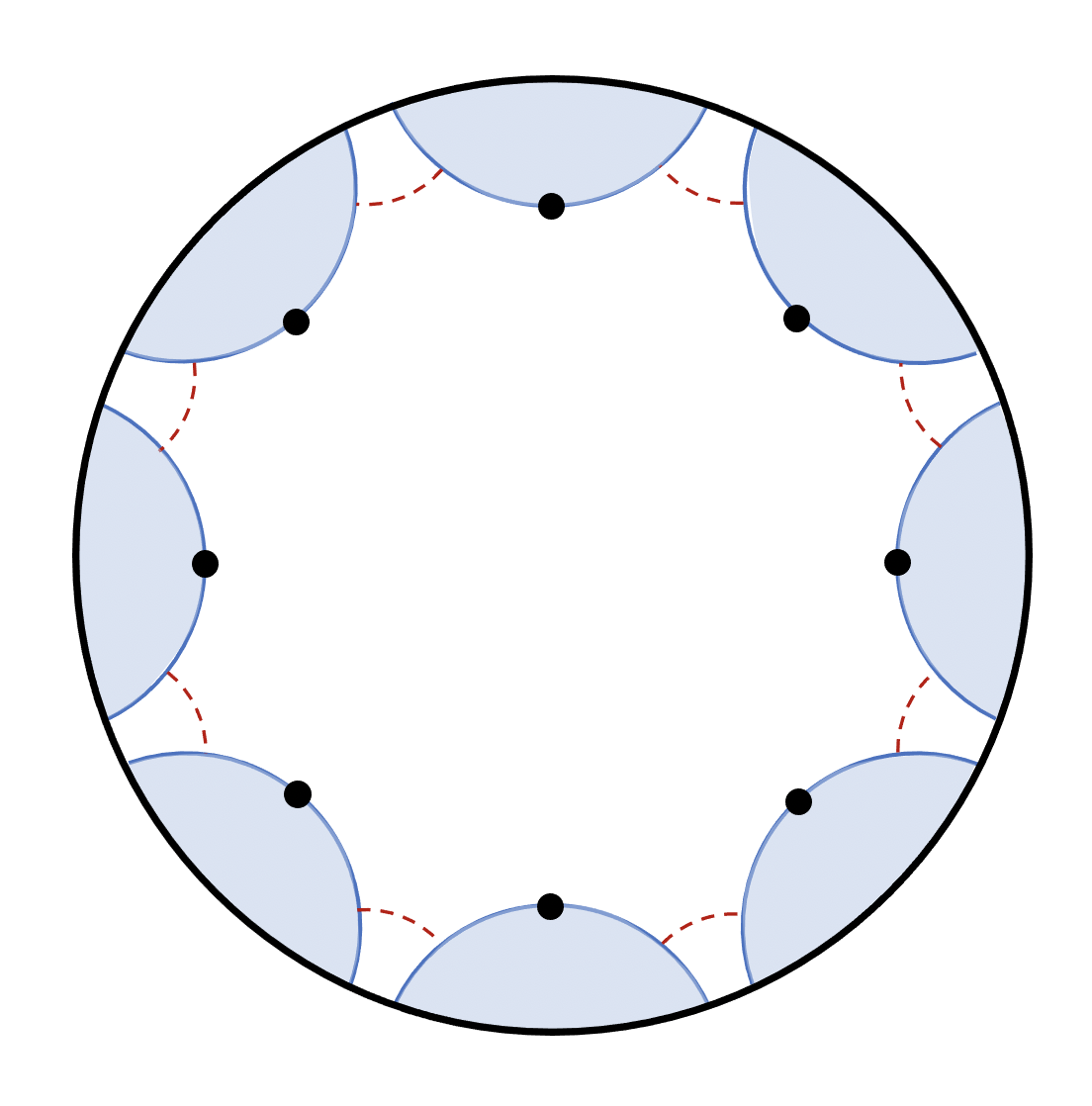}
\caption{\it An eight-defect spatial geometry at a moment of time reflection symmetry, when all defects are at rest.  The circular arcs extending to the conformal boundary on either side of a given black dot are identified to make each conical defect of deficit angle $\pi$.  The red dashed line is the apparent horizon; the defects at the moment of time symmetry have reached a maximum separation that is well inside the horizon.  }
\label{fig:circus}
\end{figure}

Time-symmetric configurations where the conical defects lie fully inside an apparent horizon arise when we use $\bZ_{\sfk}$ defects with $\sfk>2$ as in~\ref{fig:Z3cutout}, or when we have more than a pair of $\bZ_2$ defects.  For example, one can consider a $\bZ_p$ symmetric array of $\bZ_2$ defects on a surface of time symmetry where they are instantaneously all at rest.  Such a configuration is depicted in figure~\ref{fig:circus} for $p=8$.

%
%

There is of course nothing sacrosanct about such a circularly symmetric array; there are various moduli corresponding to moving the locations of the defects, subject to the constraint that there is an asymptotically $AdS_3$ boundary (if the defects are too dense, one ``overcloses'' the universe and has a big bang/big crunch cosmology with a compact spatial geometry).

\subsection{Orbiting defects and rotating black holes} 
\label{sec:rotating}

Rotating black holes are obtained when the defects carry $AdS_3$ angular momentum.  This can be achieved either by generalizing the defects from the 1/2-BPS case of section~\ref{sec:STembed} to the 1/4-BPS or non-supersymmetric examples of section~\ref{sec:generalembed}; alternatively one can coherently spin up a system of multiple 1/2-BPS defects by applying independent left and right boosts to each as in~\eqref{conjugated}.

For the latter choice, there is no longer a moment of time reflection symmetry for the defect trajectories in the $\bZ_p$ symmetric array, due to the rotation.  There is however still a moment when the defects reach a maximum radius; at that point, they still have an angular velocity, and that breaks the time reversal symmetry.  

From~\eqref{charges}, the energy and angular momentum of a rotating defect geodesic with left/right boosts $\eta_{L,R}^{~}= \eta_v\pm \eta_a$ are
\be
\label{AdSenmom}
\half\big( J^3_\sl+\bar J^3_\sl\big) = N\nu \cosh \eta_v^{~} \cosh\eta_a^{~}
\quad,\qquad
\half\big( J^3_\sl-\bar J^3_\sl\big) = N\nu \sinh \eta_v^{~} \sinh\eta_a^{~}  ~.
\ee
The geodesic travels a path having
\be
\sinh^2 \rho = \cos^2(\nu\xi) \sinh^2 \!\eta_v^{~} + \sin^2(\nu\xi) \sinh^2 \!\eta_a^{~} ~,
\ee
in other words between a minimum radius $\eta_a^{~}$ and a maximum radius $\eta_v^{~}$.  Circular orbits have $\eta_v^{~}=\pm\eta_a^{~}$ (\ie\ one of $\etaL,\etaR$ vanishes).

\begin{figure}[ht]
\centering
  \begin{subfigure}[b]{0.3\textwidth}
  \hskip 0cm
    \includegraphics[width=\textwidth]{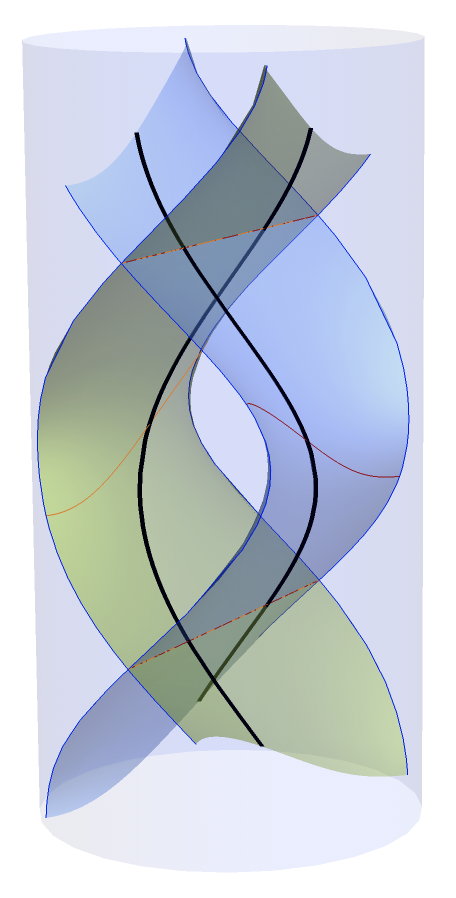}
    \caption{ }
    \label{fig:OrbDef1}
  \end{subfigure}
\qquad\qquad
  \begin{subfigure}[b]{0.3\textwidth}
      \hskip 0cm
    \includegraphics[width=\textwidth]{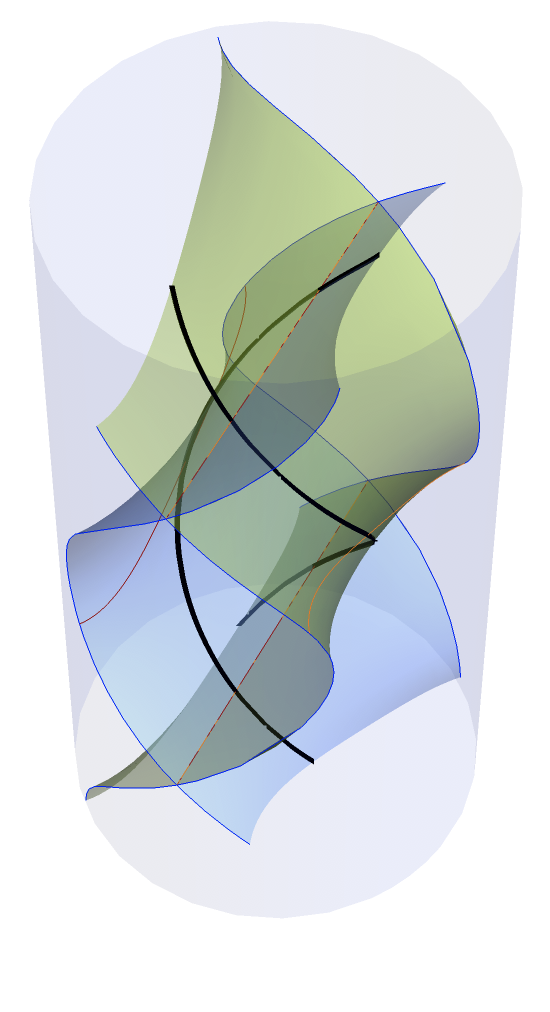}
    \caption{ }
    \label{fig:OrbDef2}
  \end{subfigure}
\caption{\it 
Two perspectives on a pair of $\bZ_2$ defects orbiting one another.  
Arcs being identified under the boosted $\pi$ rotation are indicated in red for the blue-shaded defect and orange for the green-shaded defect.   
Because of the tilt of the arcs, the region where the wedges collide is not a singularity, but rather a locus of closed timelike curves.
}
\label{fig:OrbitingDefects}
\end{figure}

Once again, we can consider a pair of defects orbiting one another.  The identifications~\eqref{combined} that make two defects of deficit angle $\alpha$ are
\begin{align}
\begin{split}
\label{2Z2rot}
\gamma_{L,1}^{~} = e^{-\etaL\sigma_1/2} \, e^{+i\alpha\sigma_3/2} \,e^{+\etaL\sigma_1/2}
~~&,~~~~
\gamma_{R,1}^{~} = e^{-\etaR\sigma_1/2} \,e^{+i\alpha\sigma_3/2} \, e^{+\etaR\sigma_1/2}
\\[.2cm]
\gamma_{L,2}^{~} = e^{+\etaL\sigma_1/2} \, e^{+i\alpha\sigma_3/2} \,e^{-\etaL\sigma_1/2}
~~&,~~~~
\gamma_{R,2}^{~} = e^{+\etaR\sigma_1/2} \,e^{+i\alpha\sigma_3/2}  \, e^{-\etaR\sigma_1/2}
\end{split}
\end{align}
and the conjugacy classes of the holonomies along a curve that surrounds both defects are
\begin{align}
\begin{split}
\big| \tr( \gamma_{L,1}^{~} \,\gamma_{L,2}^{~} )\big| &= 2(\cos\alpha\,\cosh^2\!\etaL-\sinh^2\!\etaL)
\\[.2cm]
\big| \tr( \gamma_{R,2}^{~} \,\gamma_{R,1}^{~} )\big| &= 2(\cos\alpha\,\cosh^2\!\etaR-\sinh^2\!\etaR) ~.
\end{split}
\end{align}
so that for $\bZ_2$ defects we identify
\be
2\etaL = \pi(r_+ + r_-)
~~,~~~~
2\etaR = \pi(r_+ - r_-)
\ee
The corresponding geometry is depicted in figure~\ref{fig:OrbitingDefects} 
%
%
%

There is no longer a point of collision of the two defects as they orbit around one another.  Instead, as they approach, the defects are relatively boosted, and the points being identified by a given defect are tilted by the boost so that one is later and one is earlier in the global time coordinate.  For a single boosted defect, there are no closed timelike curves under the identification, since one is just boosting a static identification under a spatial rotation.  But with multiple, relatively boosted defects, it can happen that in the set of points identified under the boosted rotations and their products, one can find regions of closed timelike curves.  These regions are separated from the region where all closed paths under the identifications are spacelike by a locus of null identifications; this latter locus is the ``singularity'' of the rotating BTZ geometry~-- not a curvature singularity, but rather a breakdown of causality (though an infinitesimal perturbation leads to a curvature singularity~\rcite{Horowitz:2002mw}).  

When defects are moving purely radially, or with sufficiently small amounts of orbital angular velocity, the locus of closed timelike curves lies inside the event horizon, and the boundary of this region is usually taken to be the singularity of the rotating BTZ black hole.  However, more serious pathologies can occur with the large deficit angles associated to global orbifolds~\rcite{DeDeo:2002yg}, in which every point in $AdS_3$ lies on a closed timelike curve.  Consider for instance the pair of orbiting defects in figure~\ref{fig:OrbitingDefects}.  The colored lines, which indicate the locus of points being identified at a particular proper time along the defect trajectory, are tilted so that one is earlier than the other; see also figure~\ref{fig:DefectRot}.  

Now consider a null geodesic along the conformal boundary traveling in the opposite sense to the defects' rotation.  Every time the geodesic passes through a wedge of identification, it gets shifted backward in time.  It was shown in~\rcite{DeDeo:2002yg} that for circular orbits (\eg\ when $\etaR=0$ for both defects), when the sum of the two defect angles equals $2\pi$, this null curve closes on itself; and when the sum of the angles exceeds $2\pi$, there are closed timelike curves passing through every point on the conformal boundary.  Thus the entire spacetime is pathological.

What this result means is that there is a limit to the angular velocity of the defects which avoids this pathology.  For instance, the pair of $\bZ_2$ defects in figure~\ref{fig:OrbitingDefects} must have $\eta_{R,i}^{~}\ne0$ so that they are not quite on circular orbits.  For $\bZ_\sfk$ defects with $\sfk>2$, one cannot achieve a circular orbit without encountering closed timelike curves at much smaller values of the axial boost.

The physical spacetime is usually taken to be the region outside the locus of closed null/timelike curves, and the region having closed timelike curves is excised.  Note that from the point of view of the worldsheet sigma model, this is not allowed under the rules of constructing orbifolds~-- one must consider the entire group manifold and carry out the quotient by whatever discrete subgroup is being gauged.  Perturbative string theory breaks down at the singularity, and one needs the a complete non-perturbative description in order to regularize the singularity and follow the subsequent evolution.

In general, it is complicated to determine the location of the singularity analytically.  But when the locus of null identifications reaches the conformal boundary, it has a fixed point.  For example, for two $\bZ_2$ defects the fixed points are located at $\tau=\pm\pi/2$, $\sigma=\pm\pi/2$, just as they were for non-orbiting defects, and one recovers the points in figure~\ref{fig:OrbDef1} where the singularity hits the boundary.

Another option for making systems with $AdS_3$ angular momentum is to consider the general defects~\eqref{orbact} which for $\sfm,\sfn$ both nonzero carry $AdS_3$ as well as $\bS^3$ angular momentum 
\be
h-\bar h = N\sfm\sfn /\sfk^2
~~,~~~~
J_\cR-\bar J_\cR= N\sfm/\sfk~.
\ee  
We can again take a single such defect and boost it radially as we did above for 1/2-BPS defects to displace it radially on the surface of time symmetry; and again make a $\bZ_p$ symmetric array which now carries $AdS_3$ angular momentum because the individual defects do.


\section{Euclidean continuation: Conical defects in $\bf \hthree$} 
\label{sec:euclidean}

Because the Lorentzian orbifold that describes colliding/orbiting defects has unphysical regions of closed timelike curves, it is highly unlikely that the entire orbifold geometry survives a proper treatment in non-perturbative string theory.  While an investigation of the nature of the singularity is needless to say of great interest, we will need more sophisticated tools to do so.  The worldsheet formalism is on firmer ground if we can avoid such pathologies, and indeed we can by passing to the Euclidean theory.  Here the defects typically avoid one another, and there is a self-consistent perturbative expansion around the orbifold background.


\subsection{Euclidean continuation of $\it AdS_3$\,: the hyperbolic ball $\bf\hthree$}
\label{sec:hthree}

While everywhere we refer to the $AdS_3$ isometries in terms of the group $\sltwo$, it is somewhat more convenient to use an equivalent parametrization of $AdS_3$ as the $SU(1,1)$ group manifold, as it diagonalizes the global time translation isometry and thus simplifies the analytic continuation to the Euclidean section.  Elements of $SU(1,1)$ can be written as
\be
\label{sl2mat}
\xx = \bigg( \begin{matrix} \,t+iz &~ x-iy\,\, \\ \, x+iy &~ t-iz\,\, \end{matrix} \bigg)
~~,~~~~
\det\big[ \xx \big] = t^2+z^2-x^2-y^2 = 1~.
\ee
Without the determinant constraint, one has $\bR^{2,2}$ with metric $ds^2\tight= -dt^2\tight-dz^2\tight+dx^2\tight+dy^2$.  The constraint is solved in terms of Euler angles
\be
\label{Euler}
g_\sl = e^{\frac i2(\tau-\sigma)\sigma_3} \, e^{\rho\sigma_1}\, e^{\frac i2(\tau+\sigma)\sigma_3}
= \bigg( \begin{matrix} \,e^{i\tau}\cosh\rho\; & \;e^{-i\sigma}\sinh\rho\,\, \\ \, e^{i\sigma}\sinh\rho\; & \;e^{-i\tau}\cosh\rho\,\, \end{matrix} \bigg)  ~,
\ee
and the $AdS_3$ metric in this global parametrization is the induced metric on the constraint surface
\be
ds^2 = \half \tr\big[ g^{-1}dg\, g^{-1}dg \big]  
= d\rho^2 + \sinh^2\!\rho\, d\sigma^2 - \cosh^2\!\rho\, d\tau^2  ~.
\ee

The trajectory 
\be
g_\sl = e^{i\nu\xi\sigma_3}
\ee
is a geodesic $\tau = \nu\xi$ sitting at $\rho=0$.  
Conical defects are made by considering the two surfaces $\sigma=\pm\alpha/2$, cutting out the wedge between them, and identifying the two sides under a discrete rotation by an angle $\alpha$ 
\be
g_\sl \sim e^{-i\alpha \sigma_3/2} \, g_\sl \, e^{+i\alpha\sigma_3/2}
\ee
that keeps fixed this geodesic.

Conjugation by left and right boosts as in equation~\eqref{geoboost} moves the geodesic onto a trajectory that oscillates around the center of $AdS_3$ and deforms the two geodesic surfaces that are identified to made the geometry with a boosted defect.

Euclidean $AdS_3$ is obtained by Wick rotating $z\to -iz$ so that we now parametrize the diagonal elements in terms of light cone coordinates $t\pm z$
\be
\label{h3mat}
\xx = \bigg( \begin{matrix} \,t+z &~ x-iy\,\, \\ \, x+iy &~ t-z\,\, \end{matrix} \bigg)
~~,~~~~
\det\big[ \xx \big] = t^2-z^2-x^2-y^2 = 1~;
\ee
the space of these matrices without the constraint is simply $\bR^{3,1}$, with $\hthree$ the hyperboloid of timelike, future-directed vectors of fixed length, preserved by Lorentz transformations $SO(3,1)\simeq\sltwoc$, which we can parametrize via the Euclidean continuation $\tau = -i\tauE$ of~\eqref{Euler} 
\be
\label{EucEuler}
\xx =  \bigg( \begin{matrix} \,e^{\tauE}\cosh\rho\; & \;e^{-i\sigma}\sinh\rho\,\, \\ \, e^{i\sigma}\sinh\rho\; & \;e^{-\tauE}\cosh\rho\,\, \end{matrix} \bigg)  ~;
\ee
the metric continues to
\be
ds^2 = d\rho^2 + \sinh^2\!\rho\, d\sigma^2 + \cosh^2\!\rho\, d\tauE^2  ~.
\ee
A vector in $\bR^{3,1}$ is realized here as a Weyl bispinor, and the Lorentz group acts via two commuting copies of complexified $\sutwo$ on the left and right:
\be
\xx \mapsto \gamma^\dagger\,\xx\, \gamma
~~,~~~~
\gamma = e^{i(\omega_i - i\eta^i)\sigma_i/2}  ~.
\ee
The independent left/right $\sltwo$ actions of the Lorentzian $AdS_3$ have Wick rotated into these two commuting complexified $\sutwo$ actions.
The space of the matrices $\xx$ with the constraint~\eqref{h3mat} is not a group manifold, rather it is the symmetric space $\hthree = \sltwoc/\sutwo$.  

The relation between the independent left and right $\sltwo$ actions of the Lorentzian section and the pair of complexified $\sutwo$ actions in the Euclidean section is made manifest by rewriting the left and right $\sltwo$ actions as vector and axial transformations
\begin{align}
\begin{split}
\gamma_L^{~} = e^{(i\alpha^3_L \sigma_3 + \alpha^1_L\sigma_1 + \alpha^2_L\sigma_2)/2}
~~&,~~~~ 
\alpha^i_L = \alpha_a^i+\alpha_v^i
\\[.2cm]
\gamma_R^{~} = e^{(i\alpha^3_R \sigma_3 + \alpha^1_R\sigma_1 + \alpha^2_R\sigma_2)/2}
~~&,~~~~ 
\alpha^i_R = \alpha_a^i-\alpha_v^i  ~.
\end{split}
\end{align}
Then the Wick rotation to $\hthree$ simply rotates 
\begin{align}
\begin{split}
\label{gpWick}
\alpha^{1,2}_v \leftrightarrow i\omega^{1,2}
~~&,~~~~
\alpha^3_v \leftrightarrow \omega^3
\\[.2cm]
\alpha^{1,2}_a \leftrightarrow \eta^{1,2}
~~~\,&,~~~~
\alpha^3_a \leftrightarrow  -i\eta^3~~.
\end{split}
\end{align}

Translations along $\sigma$ in the parametrization~\eqref{Euler} of Lorentzian $AdS_3$ are vector transformations by $\alpha^3_v$, which Wick rotate to the rotations along $\omega^3$ in $\hthree$; translations along $\tau$ are associated to the axial transformations $\alpha^3_a$, which Wick rotate to the boosts $\eta^3$ along $\tauE$ in $\hthree$.  In the limit $\rho\to\infty$, the standard polar coordinates $\vartheta,\varphi$ on the $\bS^2$ conformal boundary of $\hthree$ are related to $\tauE,\sigma$ via
\be
\varphi = \sigma
~~,~~~~
\cos \vartheta = \tanh \tauE  ~.
\ee


\subsection{Euclidean continuation of conical defects: orbifolds $\bf\hthree/\Gamma$}
\label{sec:hthreeorbs}

The elliptic identification under spatial rotations in $AdS_3$ that makes a static conical defect thus continues to an elliptic identification in $\hthree$~\rcite{Krasnov:2001va}.  The left and right boosts~\eqref{geoboost} that make a radially oscillating defect have $\etaL\tight=\etaR\tight\equiv\eta$ and so are axial transformations that Wick rotate to the same boost $\eta$ in $\hthree$.  Thus the transformations~\eqref{2Z2} that generate the discrete group of identifications of $AdS_3$ also define a corresponding discrete group of identifications of $\hthree$.

A simple way of characterizing geodesics, in either $\hthree$ or $AdS_3$, is to work with the unprojected matrices $\hat\xx$ in $\bR^{3,1}$ or $\bR^{2,2}$, before imposing the condition of unit determinant.  Geodesics in these Cartesian spaces are straight lines; those with nonzero $\det[\hat\xx]$ project down onto geodesics in hyperbolic space via $\xx=\hat\xx/\!\sqrt{\det[\hat\xx]}$.  We can take a straight line in $\bR^{2,2}$ that represents a geodesic in $AdS_3$, and Wick rotate it to get a corresponding geodesic in $\hthree$. 
For instance, the static geodesic at the center of $AdS_3$ is the line 
\be
\label{geostatic}
t=1
~~,~~~~
z=\xi
~~,~~~~
x=y=0  ~~.
\ee 
The same line in the Cartesian coordinates on $\bR^{3,1}$ projects down to a geodesic that runs from pole to pole of the boundary sphere running radially through $\hthree$.
Then for $\xi=0$, in $\bR^{2,2}$ we have a timelike vector at the origin in $x,y,z$ which projects down onto $\tau=0,\rho=0$ in $AdS_3$, and $\tauE=0,\rho=0$ in $\hthree$.  The surfaces $\tau=0$ in $AdS_3$ and $\tauE=0$ in $\hthree$ both have the geometry of the Poincar\'e disk where we can continue from one to the other; the above geodesics pass through the center of the disk.  

The planar surfaces being identified by a rotation are also the same in $\bR^{2,2}$ and $\bR^{3,1}$, and can be parametrized by
\be
\label{rotplanes}
t=1
~~,~~~~
z=\xi
~~,~~~~
x+iy = r\, e^{i(\phi\pm\alpha/2)}
\ee
with $\xi\in\bR,r\in \bR_+$ for a wedge centered about the direction $\phi$ and identified under a rotation by angle $\alpha$.

Similarly, the radially boosted geodesic is radially boosted in both Cartesian spaces $\bR^{2,2}$ and $\bR^{3,1}$
\be
\label{Egeoboost}
t=\cosh\eta
~~,~~~~
z=\xi
~~,~~~~
x+i y = e^{i\phi}\sinh\eta
\ee
so that at the moment of time reflection symmetry $\xi=0$, the geodesic is displaced from the origin to $\rho=\eta$ in both $AdS_3$ and $\hthree$.
The planes~\eqref{rotplanes} being identified boost to some other planes that are identified under a rotation conjugated by the boost, which are again the same on the surface of time reflection symmetry.

\begin{figure}[ht]
\centering
  \begin{subfigure}[b]{0.4\textwidth}
  \hskip 0cm
    \includegraphics[width=\textwidth]{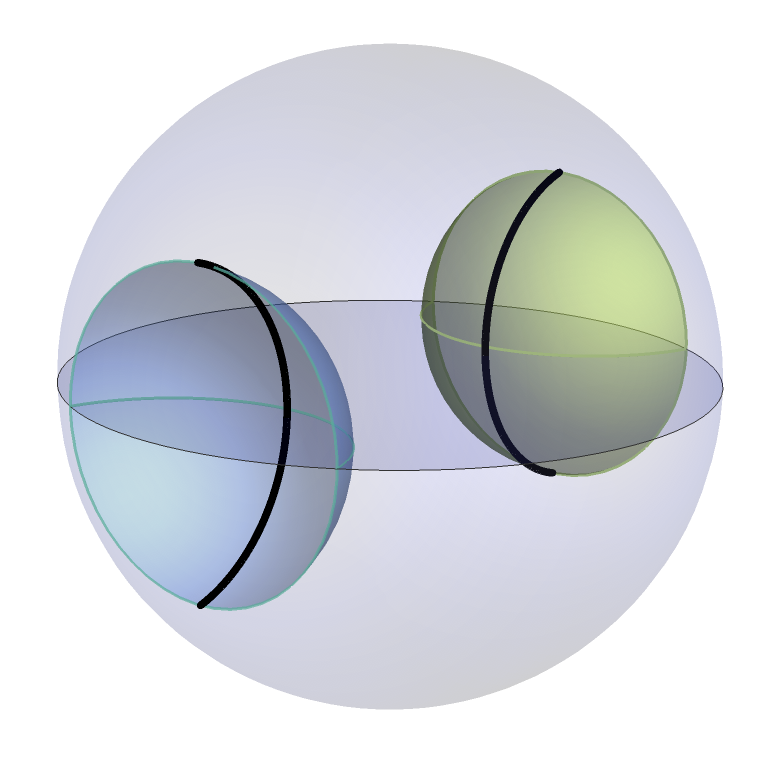}
    \caption{ }
    \label{fig:EucOrb}
  \end{subfigure}
\qquad\qquad
  \begin{subfigure}[b]{0.39\textwidth}
      \hskip 0cm
    \includegraphics[width=\textwidth]{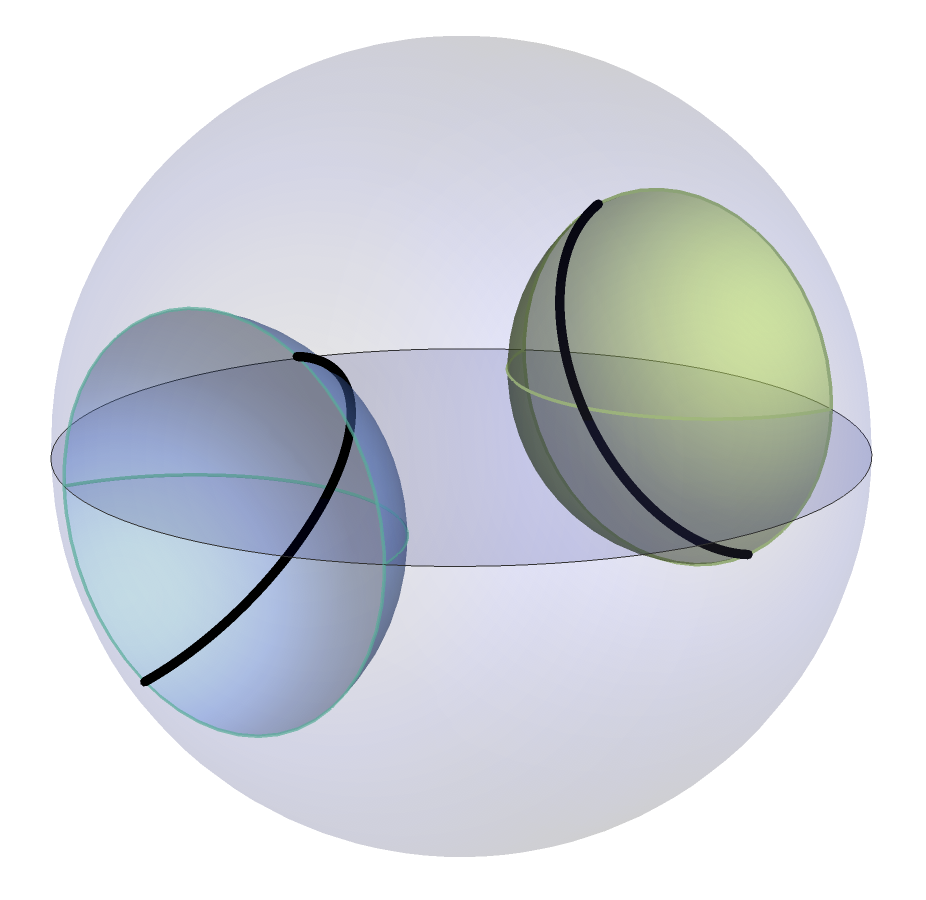}
    \caption{ }
    \label{fig:EucRotDefect}
  \end{subfigure}
\caption{\it 
On the left, the Euclidean continuation of non-orbiting defects.  The surface of time reflection symmetry continues to the equatorial Poincar\'e disk $\bH_2$; the Lorentzian particle geodesics continue to the black geodesics in $\hthree$.  A $\bZ_2$ defect is obtained by identifying the shaded spherical surfaces on either side of a given geodesic, removing the interior (the shaded region in figure~\ref{fig:cutout}).
On the right, the Euclidean continuation of orbiting defects.  
}
\label{fig:EucWedges}
\end{figure}

The Euclidean continuation of a pair of oppositely boosted $\pi$ defects (whose geometry on the surface of time reflection symmetry is depicted in figure~\ref{fig:cutout}) is shown in figure~\ref{fig:EucOrb} (see also figures~\ref{fig:EucZ2-3D} and~\ref{fig:EucZ3-3D}).  Note that the geodesics travelled by the defects are in general circles orthogonally intersecting the spherical boundary of $\hthree$, and the ``wedges'' being removed are bounded by segments of spheres between two longitude lines (in the orientation where the poles are the locations where the defect hits the boundary); these spherical segments also orthogonally intersect the boundary.
Because the defect location is at a radial extremum on both Euclidean and Lorentzian sections as a function of time, it is at its maximum radius on the Lorentzian section while it is at its minimum radius in the Euclidean section (see figures~\ref{fig:twoZ2} and~\ref{fig:DefectCollisions}).

Generalizing to nonzero angular momentum means turning on some amount of ``axial boost'' in $AdS_3$ (parametrized by $\eta_a$ in equation~\eqref{AdSenmom}) that leads to a rotation that tilts the trajectory in $\hthree$ according to~\eqref{gpWick}.  The boost in $\sltwoc$ displaces the radial extremum of the geodesic away from the origin as it did for the non-rotating defect, and the rotation (around the same axis as the boost) tilts it so that it is no longer perpendicular to the equatorial plane; see figure~\ref{fig:EucRotDefect}.  The lift of the geodesic analogous to~\eqref{Egeoboost} is a straight line, tilted away from running along the $\hat{\bf z}$ direction by the rotation:
\be
\label{rotatedline}
t=\cosh\eta_a
~~,~~~~
z = \xi\cos\omega
~~,~~~~
x+iy = e^{i\phi}\sinh\eta_a + \xi\sin\omega ~.
\ee 
This projects down in $\hthree$ to a tilted geodesic of the sort depicted in figure~\ref{fig:EucRotDefect}.
The $\bR^{2,2}$ version of this geodesic is the same straight line, parametrized as
\be
\label{boostedline}
t=\cosh\eta_a
~~,~~~~
z = \xi\cosh\eta_v
~~,~~~~
x+iy = e^{i\phi}\sinh\eta_a + \xi\sinh\eta_v ~.
\ee 

For orbiting defects, the line is tilted in a plane perpendicular to the radial direction in the unprojected Cartesian space; the tilting is a boost in $\bR^{2,2}$ and a rotation in $\bR^{3,1}$.  We demand that the lines representing the geodesic coincide and in particular have the same slope; this relates the boost rapidity $\eta_a$ in $\bR^{2,2}$ to the rotation angle $\omega$ in $\bR^{3,1}$ via
\be
\tanh \eta_v = \tan \omega
\ee

The  identification of $\hthree$ under a rotation by angle $\alpha$ is also tilted into
\be
\xx \;\sim\; \gamma^\dagger \xx \gamma
~~,~~~~
\gamma = e^{\frac i2(\omega - i\eta_a)\sigma_1}\, e^{-\frac i2 \alpha \sigma_3} \,e^{-\frac i2(\omega - i\eta_a)\sigma_1}  ~.
\ee
The boosted wedge being cut out of Lorentzian $AdS_3$, whose sides are identified by a boosted rotation that fixes the boosted defect geodesic, continues into a spherical wedge being cut out from $\hthree$, with the sides being identified by an elliptic transformation in $\sltwoc$ which fixes the circular geodesic in $\hthree$ that is the analytic continuation of the Lorentzian defect trajectory.


Note that the appropriate time coordinate for the Wick rotation can no longer be taken to be the global time on the covering space.  The problem is that there is no simple hypersurface of the Lorentzian geometry that contains all of the wedges being cut out of the full 3d geometry (the identifications involve some timelike motion due to the way that the geodesics are tilted), and no way of simply matching to a corresponding hypersurface of the Euclidean geometry with identical wedges cut out.  The appropriate parametrization of the geometry co-rotates with the defects.

\subsection{Defect correlators} 
\label{sec:correlators}

We thus have a prescription for the Euclidean continuation of a collection of conical defects.  The Euclidean geometry is an orbifold of $\hthree$ generated by a collection of elliptic elements $\{\gamma_{L,i}^{~},\gamma_{R,i}^{~} \tight= \gamma_{L,i}^{\dagger} \}$ implementing $\bZ_{\sfk_i}$ twists.

In contrast to the general 1/2-BPS state, the conical defects here are special~-- their Euclidean versions locally near the defect are $\bZ_\sfk$ quotients of $\hthree\times\bS^3$, and the bulk geometry is globally a quotient
\be
\big( \hthree\times\bS^3 \big) / \Gamma  ~,
\ee
where $\Gamma$ is the discrete subgroup of $\sltwoc$ generated by the collection of elliptic identifications for the defects.  Each conical defect travels a geodesic (circular arc) in $\hthree$, landing on the conformal $\bS^2$ boundary at a pair of conjugate points $z_i, z_i'$.  In the non-orbiting case, these two points are symmetric about the equator, $z_i'=z_i^*$, in order to have a surface of time reflection symmetry; though in general (\eg\ for the orbiting case, and more generally for generic correlators of heavy operators) one can place the fixed points of the elliptic transformations independently on the sphere.
In the dual CFT, each conical defect corresponds to a conjugate pair of local operators inserted at the corresponding points $z_i$ and $z_i'$.  

The holographic map for these operators is precisely understood.  As reviewed above, for the 1/2-BPS $\bZ_\sfk$ conical defects, these operators locally transform the vacuum state, which is a condensate of strings with the lowest unit of winding $1/n_5$, into a condensate of strings of winding $\sfk/n_5$.  The fivebranes back-react by assuming a spiral shape at the source.
In $(\cM)^N/S_N$ symmetric orbifold terms,
the $\bZ_\sfk$ conical defect is associated to a particular orbifold twist operator where all the cycles have length $\sfk$.  These operators create states in the R-R sector of the CFT, and carry spacetime quantum numbers~\eqref{qnums}.

In string theory one uses such operators, having conformal dimension a finite fraction of the central charge $\cst =6N$, to construct a ``heavy'' background~-- a state macroscopically excited away from the spacetime CFT vacuum~-- and then worldsheet string theory calculates the correlation functions of perturbative excitations around this state.  In the spacetime CFT, these are correlation functions with two ``heavy'' operator insertions as well as some number of ``light'' operator insertions (whose conformal dimension $h\ll \cst$).  Correlation functions involving a single heavy defect have been analyzed in~\rcite{Hijano:2015rla,Bufalini:2022wzu}; see~\rcite{Hijano:2015qja} for multipoint correlators.


The geometries we have built above are the leading semi-classical approximation (in $1/N$) to spacetime CFT correlation functions of several heavy operators (one conjugate pair for each defect).  Furthermore, these heavy backgrounds are exact solutions to classical string theory, non-perturbatively in~$\alpha'$.   We saw this in section~\ref{sec:nullgauging} for a single conical defect, where the single $\bZ_\sfk$ identification could be diagonalized, and the $U(1)$ factor in the CFT where the orbifold acts as a shift symmetry could be isolated.  Correlation functions of the orbifold are those of the cosets $\frac\sltwo\uone$ and $\frac\sutwo\uone$ times free field vertex operators on a modified lattice of windings and momenta resulting from the shift orbifold.

For multiple defects, the $\bZ_{\sfk_i}$ generators of the various defects are not simultaneously diagonalizable, and so one has a non-abelian orbifold group~-- an infinite discrete subgroup of $\sltwoc\times \sutwo_L\times\sutwo_R$ known as a {\it Kleinian group}.  If we consider orbiting, 1/4-BPS, or non-supersymmetric conical defects, the orbifold is asymmetric.

To compute worldsheet amplitudes in the presence of multiple defects, we might look for a suitable generalization of the methods developed in~\rcite{Kutasov:1999xu,Maldacena:2000kv,Maldacena:2001km,Teschner:1999ug,Teschner:2001gi,Ponsot:2002cp,Hikida:2007tq,Dei:2021xgh,Dei:2021yom,Dei:2022pkr,Bufalini:2022toj,Ashok:2020dnc,Nippanikar:2021skr,Ashok:2022vdz} in order to build string vertex operators invariant under the group $\Gamma$.  There, a central role is played by the coherent state vertex operator $\Phi_j(x,\bar x)$ built from a somewhat different matrix parametrization of $\hthree$
\be
\label{hmat}
\sfM = 
\left(\begin{matrix} 1 & 0 \\ \chi & 1 \end{matrix}\right)
\left(\begin{matrix}
e^\phi & 0 \\
0 & e^{-\phi}  
\end{matrix}\right)
\left(\begin{matrix} 1 & \bar\chi \\ 0 & 1 \end{matrix}\right)
=
\left(\begin{matrix}
e^\phi & e^\phi \bar\chi \\
e^\phi \chi & ~e^{-\phi} + e^{\phi} \chi\bar\chi
\end{matrix}\right)~,
\ee
on which $\gamma\in\sltwoc$ again acts via $\sfM\to \gamma^\dagger\sfM \gamma$.  The functions
\be
\label{scalingfns}
\Phi_j(x,\bar x) = \frac{2j-1}{\pi}\biggl(\bigl(x,1)\cdot \sfM\cdot \biggl(\begin{matrix} \bar x \\ 1\end{matrix}\biggr)\biggl)^{-2j}
= \frac{2j-1}{\pi} \Big( |\chi-x|^2e^{\phi} + e^{-\phi} \Big)^{-2j}
\ee
are eigenfunctions of the Laplacian on $\bH_3^+$. The complex parameter $x$ labels points on the boundary. The operator $\Phi_j(x,\bar x)$ transforms as a tensor of weight $(j,j)$ under $\sltwoc$.  Formally, one can construct untwisted sector vertex operators on the orbifold by summing this operator over its images under the orbifold group $\gamma(x), \gamma\in\Gamma$, appropriately regulating the infinite sum involved if necessary.

As shown in~\rcite{Martinec:2018nco,Martinec:2022okx}, for a single defect there are untwisted sector vertex operators describing supergravity excitations around the defect; for the multi-defect geometries with sufficiently sharp conical defects (thus having large redshift to the orbifold point), or sufficiently well-separated defects, the untwisted sector operators will be fairly well localized around an individual defect, with tails that overlap the other defects.  As usual for hyperbolic manifolds, the spectrum of the Laplacian will be chaotic.

There will be in addition twisted sector vertex operators that deform the winding condensate associated to the defect.  One expects such winding operators for each defect when several are present, and more generally for each element $\gamma\in\Gamma$; there are winding strings that only close up to the action of $\gamma$.  It would be interesting to understand their properties, as they describe perturbations of the Coulomb branch tails of the little string condensate carried by the underlying fivebranes.

\begin{figure}[ht]
\centering
\includegraphics[width=0.3\textwidth]{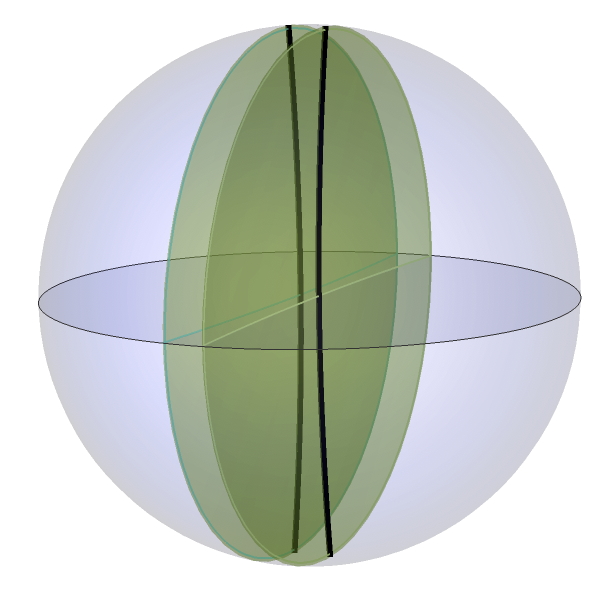}
\caption{\it The OPE limit of the correlator of two $\bZ_2$ defects.  }
\label{fig:OPElimit}
\end{figure}

Of particular interest is the pair of $\bZ_2$ defects discussed above.  As we turn off the boost $\eta$ that separates their insertion points on the conformal boundary, we approach the OPE limit of the correlator, see figure~\ref{fig:OPElimit}.  
%
%
The leading term in the OPE of two 1/2-BPS $\bZ_2$ R-R defect operators with $h\tight=\bar h\tight=N/4$ and $J\tight=\bar J\tight=N/4$ is a 1/2-BPS NS-NS defect operator with $h\tight=\bar h\tight=N/2$ and $J\tight=\bar J\tight=N/2$; the associated state  is an extremal BTZ black hole.  Indeed, in the construction of section~\ref{sec:conjugation}, the limit where we send the relative boost $\eta$ to zero, the two defects touch; the closed geodesic on the surface of time symmetry (the red dashed line in figure~\ref{fig:cutout}) shrinks to zero size and the area of the apparent horizon vanishes.%
\footnote{This may be one instance where the possibility of conjugation in $\sutwo$ mentioned briefly in section~\ref{sec:conjugation} could be of use~-- we can regularize the OPE in $\hthree$ by separating the two $\bZ_2$ defect rings in $\bS^3$.  }

An intriguing weak-coupling counterpart to this process involves the operator product of the corresponding defect operators in the symmetric orbifold.  These are labelled by the conjugacy class of the symmetric group consisting of $N/2$ disjoint transpositions (2-cycles), which we denote $(2)^{N/2}$.  The product of two such twists decomposes on a variety of conjugacy classes, but at large $N$ these are all concentrated on classes consisting of a handful of cycles whose lengths are of order $N$ together with a smattering of shorter cycles.  

Heuristically, the reason is as follows.  Each twist operator is a sum over all permutations within the conjugacy class $(2)^{N/2}$.  Their product thus contains many terms; consider some random term in the sum.  Using an overall relabelling of letters, we can take the first permutation to be $P_1=(12)(34)\dots (N-1,N)$, and the second to be $P_2=(\sigma_1\sigma_2)(\sigma_3\sigma_4)\dots(\sigma_{N-1}\sigma_N)$.  At the first step, we take the first transposition $(12)$ of $P_1$ and ask where $P_2$ sends the letter $2$.  It is either $1$, and the cycle length in the product is one, or it is some other letter $\ell$; since there are $N-2$ chances that it is not $1$, it is vastly more likely that the cycle length is at least two.  One then asks where $P_1$ sends $\ell$, call it $\ell'$, and one asks where $P_2$ sends $\ell'$; again it is either $1$ or (much more likely) some other letter $\ell''$, and so on until one eventually closes the cycle.  The odds are that one will have to wait some finite fraction of the $(N/2)$ transpositions until one manages to hit on the one that closes the cycle.  

Once one has closed the cycle, it will generically be before we have run through all the transpositions in $P_1$, so we begin anew constructing another independent cycle in the product; and so on, until one has used up all the letters.  Thus while there will generically be a handful of long cycles taking up most of the letters, once one is down to the last few transpositions in this process, the cycles are short and this is why the generic element in the product has a mix of both long and short cycles.

\begin{figure}[ht]
\centering
  \begin{subfigure}[b]{0.4\textwidth}
  \hskip 0cm
    \includegraphics[width=\textwidth]{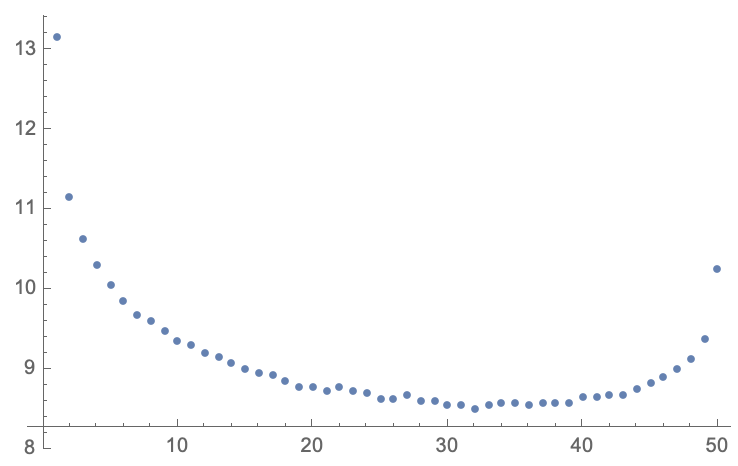}
    \caption{ }
    \label{fig:dist1}
  \end{subfigure}
\qquad\qquad
  \begin{subfigure}[b]{0.39\textwidth}
      \hskip 0cm
    \includegraphics[width=\textwidth]{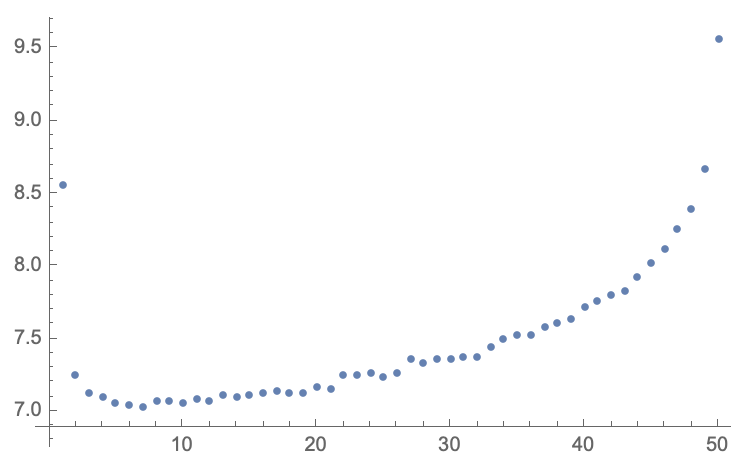}
    \caption{ }
    \label{fig:dist2}
  \end{subfigure}
\caption{\it Cycle length distribution in random sampling containing $10^5$ permutations in the conjugacy class of the symmetric group associated to $\bZ_2$ defects in the symmetric orbifold CFT $(\cM)^N/S_N$.  The horizontal axis of each plot is the percentage $p$ of the total $N$ belonging to the cycle in question.  This never exceeds 50\% for reasons explained in the text.
On the left, a plot of the log of the number $f$ of times cycles of a given length appear in the $10^5$ sample products.
On the right, the distributions of the log of $p\tight\times \!f$, \ie\ the relative likelihood that a given copy of $\cM$ is in a cycle of length $p$.
}
\label{fig:EucWedges}
\end{figure}

Indeed, a random sampling of $10^5$ examples of permutations in the class $(2)^{N/2}$ (with $N=10^4$) yields a statistical distribution of cycle lengths for their product (shown in figure~\ref{fig:dist1}), and distribution weighted by the number of copies of $\cM$ in the cycle (shown in figure~\ref{fig:dist2}).  In this example, cycle lengths in the product always come in pairs~-- the cycle in the product containing the letter $2\ell-1$ in $P_1$ and the cycle in the product containing the letter $2\ell$ in $P_1$ are always disjoint and have the same length.  This is why in the figure, the cycle lengths cannot exceed $N/2$.

This is just what one would expect for the CFT state describing an extremal black hole~-- the resulting state is in the Hagedorn phase of the little string.%
\footnote{Which in the case of the symmetric orbifold is the same as the fundamental string, since this CFT lies in the cusp of the moduli space that describes $n_1=N$ fundamental strings bound to a single fivebrane $n_5=1$~\rcite{Seiberg:1999xz,Larsen:1999uk}, and thus there is no fractionation phenomenon.  There are multiple cusps of the moduli space, one for each factorization of $N$ into $(n_1,n_5)$.  The cusps with $n_5>1$ are in the strong-coupling regime of the spacetime CFT.}  Since the OPE is BPS at leading order in the separation of the operators (and therefore not renormalized~\rcite{deBoer:2008ss,Baggio:2012rr}), this picture is robust as we go from weak to strong coupling in the CFT.%
\footnote{In the family of supertube backgrounds, one can approach the black hole threshold from below as one moves along the configuration space of 1/2-BPS supertubes at the BPS bound toward the point where it intersects the extremal black hole threshold~\rcite{Martinec:2019wzw}; here one is approaching the extremal black hole threshold from above as one merges the two $\bZ_2$ defects.}


\section{Discussion} 
\label{sec:discussion}

\subsection{Generalizations} 
\label{sec:generalizations}

While we used $\bZ_\sfk$ orbifold defects because of their exact worldsheet description in perturbative string theory, in principle one might hope to construct the supergravity solution for the Euclidean geometry with defects corresponding to any of the Lunin-Mathur geometries as local sources.  These would be multicenter ``bubbled geometries'' where each center is locally some 1/2-BPS wiggly fivebrane source whose back-reaction in supergravity was worked out in~\rcite{Lunin:2001fv} for a single source. 
Again, the relative boost of the sources breaks supersymmetry mildly, but one might hope that some of the techniques recently developed for constructing non-BPS supergravity solutions in $AdS_3$~\rcite{Ganchev:2021pgs} might be of use.  

The 1/2-BPS geometries have a spread of asymptotic deficit angles ranging from zero (global $AdS_3$) to $2\pi$ (the extremal black hole), controlled by the shape of the profile functions $\F^I(\tilde v)$.
These geometries are not pure conical defects however, except for the $\bZ_\sfk$ examples discussed above~\rcite{Lunin:2002iz}, instead there are typically long-range tails in the metric and other supergravity fields; but one can for instance describe giant graviton states and shockwaves in $AdS_3$~\rcite{Lunin:2002bj,Lunin:2002iz,Lunin:2002fw,Chakrabarty:2021sff}, and it would be interesting to investigate their collisions given the extensive holographic dictionary regarding these states that has been built up over the years (see~\rcite{Bena:2022rna} for a review).

The geometries are asymptotically conical around each 1/2-BPS state, however, and one may use the metric holonomies to characterize the results of their collisions.  For sufficiently light 1/2-BPS states with not too much kinetic energy, the collision simply makes a state with slightly larger asymptotic conical defect, as the product of the corresponding group elements remains in the elliptic conjugacy class~\rcite{Matschull:1998rv,Holst:1999tc,Birmingham:1999yt,Krasnov:2002rn,Brill:2007zq,Lindgren:2015fum}.  Eventually the state settles down and thermalizes.  One can tune the final state defect to be near the black hole threshold, either slightly below or slightly above, by tuning the asymptotic deficit angle and the relative radial boost of the defects; it would be interesting to see what phenomena arise in the bulk that characterize the Hawking-Page transition to the black hole phase.  Arguments were given in~\rcite{Martinec:2019wzw} that in the F1-NS5 system, this transition deconfines the non-abelian excitations of the little string; one might hope to find evidence for this phenomenon through the study of defect collisions, and in particular the OPE limit discussed above in the collision of two $\bZ_2$ defects to make a near-extremal BTZ black hole.  The rotating BTZ threshold, where the energy and $AdS_3$ angular momentum become equal, is also of interest.  Here the defects don't collide; they can remain macroscopically separated, but generate a null singularity inside an apparent horizon.

Another direction would be to generalize from 1/2-BPS supertube defects to 1/4-BPS superstrata traveling approximately along geodesics in $\hthree$.  In particular, there are superstrata with long $AdS_2$ throats~\rcite{Bena:2017xbt}; it would be interesting to investigate their collisions within a larger asymptotically $AdS_3$ arena, watching the throats coalesce and form an $AdS_2$ black hole from some nonsingular initial data.  Mergers of similar throats has been discussed in the BPS and near-BPS context in~\rcite{Bena:2006kb,Martinec:2015pfa}.

\subsection{The singularity} 
\label{sec:singularities}

The weak-coupling picture of the $\bZ_2$ defect collision to make a near-extremal BTZ black hole presented above seems to hold lessons more generally.  In this regime, the defects are associated to twist operators in the conjugacy class $(2)^{N/2}$ in the symmetric orbifold; the Euclidean continuation of their near-extremal collision is the OPE limit of the four-point function where the two R-R defect operators make a near-BPS pure state in the NS-NS sector that is (a) at the black hole threshold, and (b) concentrated on Hagedorn-like conjugacy classes in the symmetric product.

In the string theory description of $n_5$ NS5-branes, the effect of 1/2-BPS string vertex operators is to change the winding condensate carried by the fivebranes as in equation~\eqref{Vtransition}~\rcite{Martinec:2020gkv,Martinec:2022okx}.  Exponentiating the vertex operator $\cV^{++}_{j'=1/2,w_y=0}$ turns the $AdS_3$ vacuum into the $\bZ_2$ orbifold defect, as we see from the identification of the coherent profile~\eqref{Zkprofile} it produces (with $\sfk=2$) in the Lunin-Mathur construction.  Since the BPS OPE is a protected quantity~\rcite{deBoer:2008ss}, the above numerical evaluation of the OPE is an accurate picture of what is happening at strong coupling.%
\footnote{It would be interesting to see if some of the methods of~\rcite{Lin:2022rzw,Lin:2022zxd} could be adapted to the present context.}  
We see that the collision of two such string condensates makes a Hagedorn string gas at the black hole threshold.  Though the description of this process in the bulk is of course beyond string perturbation theory, it is a direct extrapolation of perturbative processes where we see several shorter strings combine to make longer strings as in~\eqref{Vtransition}.

As we move away from the BPS limit and above the BTZ black hole threshold (keeping the $AdS_3$ angular momentum equal to zero), the defects move apart on the surface of time reflection symmetry in $\hthree$ and $AdS_3$; they then collide more violently in $AdS_3$ at a somewhat later time, starting from rest at this larger separation.  It is tempting to regard the singularity as the locus where the string condensates associated to the defects collide and rearrange to make a more excited Hagedorn gas of little strings than the one that arises at extremality.  
This is just the sort of effect that one expects to resolve the singularity in string theory.

A major question is whether there are precursors of this collision already at the scale of the apparent horizon, which should then also be visible in the Euclidean correlators directly (since the apparent horizon is a feature of the geometries of both $\hthree$ and $AdS_3$ at the surface of time symmetry where they agree).  This is the question of where the transition to the Hagedorn phase is happening in the bulk (to the extent that it can be localized).  The phenomenon of longitudinal spreading of perturbative strings in high-energy scattering is well-known~\rcite{Gross:1987kza,Gross:1987ar,Amati:1987wq,Amati:1988tn,Susskind:1993aa,Lowe:1995ac,Polchinski:1995ta,Giddings:2007bw,Dodelson:2015uoa,Dodelson:2015toa,Dodelson:2017hyu}.  One way of phrasing the question here is whether a similar phenomenon governs little string scattering.

It might also be interesting to revisit the calculation of scattering in Lorentzian orbifolds~\rcite{Liu:2002ft,Liu:2002kb} to see the effects of winding sectors, the spread of the string wavefunction, and so on.  Even though the geometries aren't fully consistent as string backgrounds, they may contain clues about the non-perturbative dynamics.  Furthermore, one can study the Euclidean theory, which is perfectly regular, and its properties under analytic continuation to the Lorentzian section.

The fluctuations of 1/2-BPS F1-NS5 configurations have been studied in~\rcite{Alday:2006nd,Raju:2018xue}, by quantizing the restricted phase space of the 1/2-BPS supergravity solutions themselves using the parametrization in terms of the profile functions $\F^I(\tilde v)$ of section~\ref{sec:holomap}.  It was found that the fluctuations are larger than would be expected on the basis of considerations of an extremal black hole with a ``stretched horizon''.  While the ensemble includes only a tiny subset of the full fivebrane degrees of freedom, and is working in a regime below the black hole threshold where non-abelian fivebrane dynamics hasn't fully set in, these results may be an indication that indeed the extent of the fivebrane wavefunction exceeds what one might have expected on the basis of properties of the classical geometry such as local curvature invariants, \etc.

Orbiting defects generically do not collide, at least initially; instead, the orbifold identification goes null, and then timelike in the classical solution.  Small perturbations lead to singularities~\rcite{Liu:2002kb,Horowitz:2002mw}.    
Three-charge extremal black holes can be constructed using two $\bZ_2$ defects as in section~\ref{sec:rotating}, and performing mostly \eg\ left-moving boosts~$\etaL$, sending $\etaR\to 0^+$.  This separates them to radius $\rho=\half\etaL$, and then boosts them in the $AdS_3$ azimuthal direction by an amount approaching $\half\etaL$, so that the orbit becomes circular.  On the CFT side, we are coherently exciting the two defect operators with an exponential of $L_{-1}$ but not of $\bar L_{-1}$.  At $\tau=0$, the two defects lie on the horizon, as one can see from figure~\ref{fig:OrbDef1}.  Again, these geometries are close to BPS, and describe a pure state in the ensemble of such black holes.  It would be interesting to investigate the properties of the CFT operator product expansion in this limit, and see how properties of the bulk geometry are reflected in its structure.

\subsection{Dustballs and fuzzballs} 
\label{sec:dustballs}

\begin{figure}[ht]
\centering
  \begin{subfigure}[b]{0.4\textwidth}
    \includegraphics[width=\textwidth]{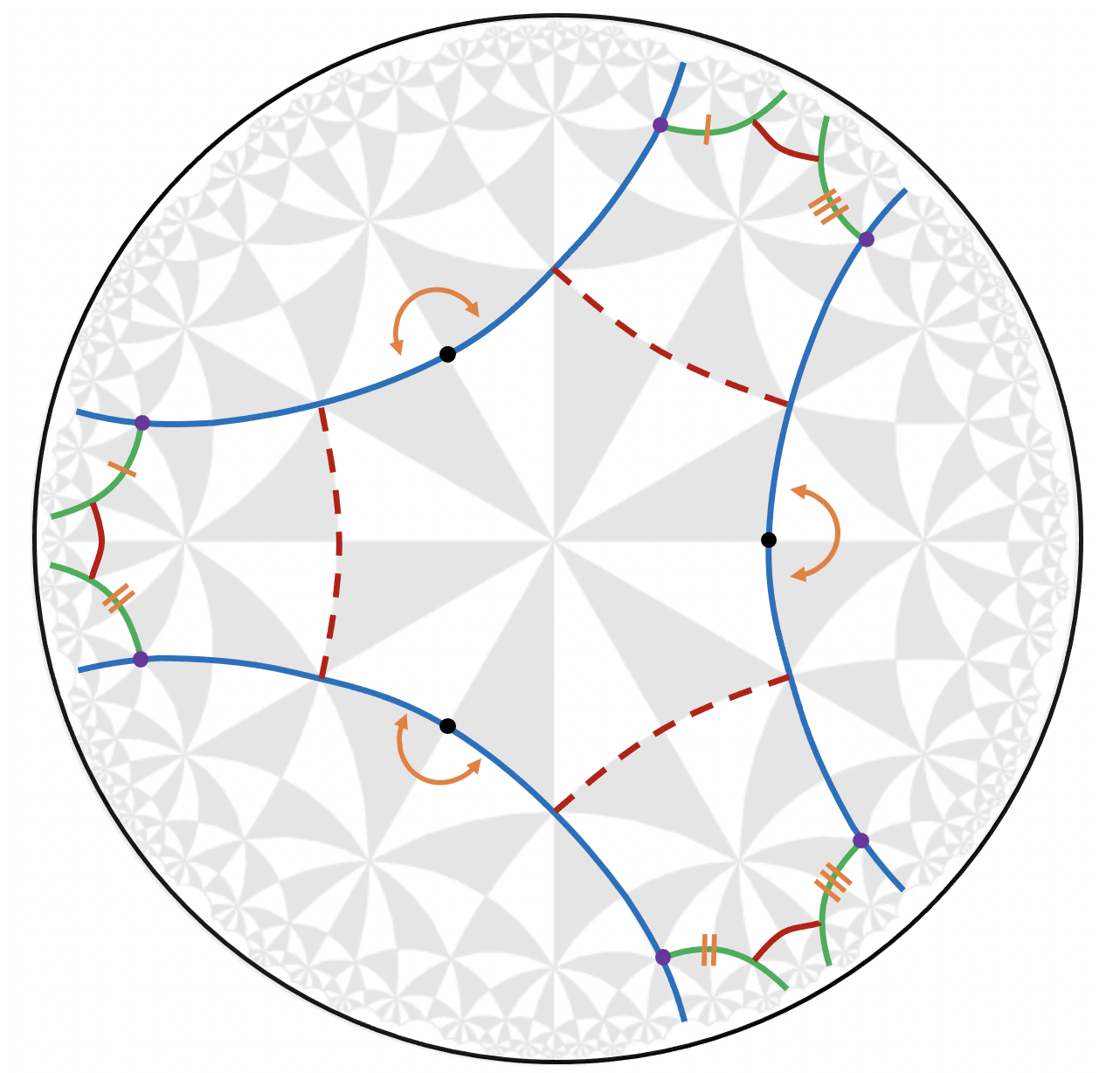}
    \caption{ }
    \label{fig:DoubleWedge}
  \end{subfigure}
\qquad\qquad
  \begin{subfigure}[b]{0.45\textwidth}
      \hskip 0cm
    \includegraphics[width=\textwidth]{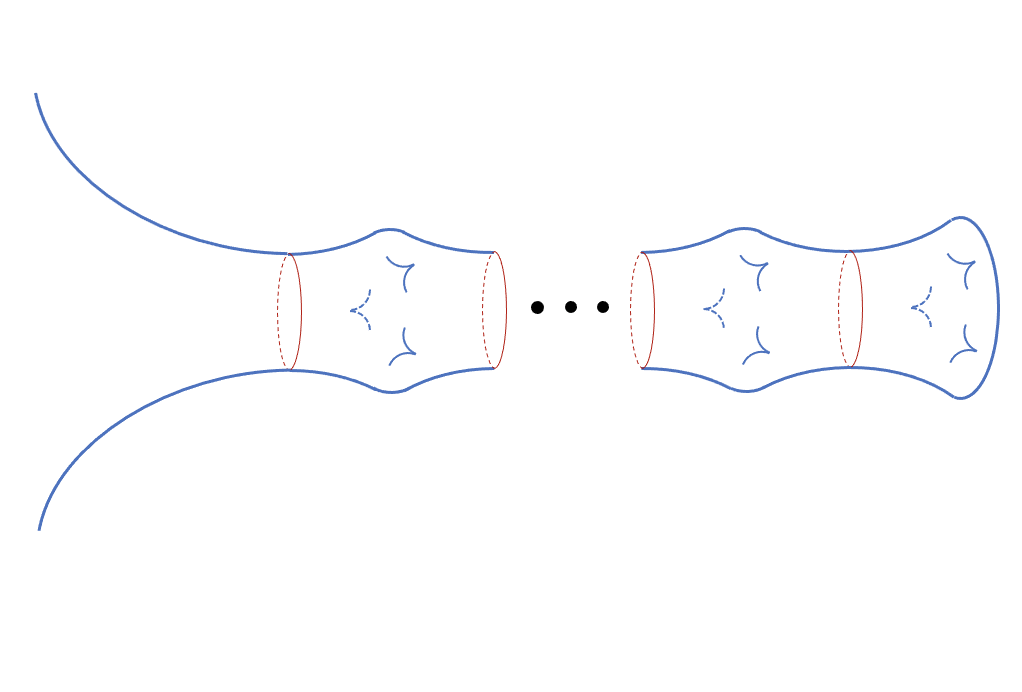}
    \caption{ }
    \label{fig:BagofGold}
  \end{subfigure}
\caption{\it 
One can concentrically place rings of defects to make a black hole with an arbitrarily long ``bag of gold" geometry behind the horizon.  On the left, the orange arrows and hash marks indicate the circular arcs to be identified to make a defect geometry on its surface of time reflection symmetry, with two rings of three $\bZ_2$ defects shown.
On the right, a cartoon of the geometry with an arbitrary number of defect rings, after the identifications.
}
\label{fig:EucWedges}
\end{figure}

One can construct arbitrarily long ``bag of gold'' spatial geometries by considering concentric rings of defects.  Basically, the defects insert positive curvature in the geometry, trying to close it off, while the negative vacuum curvature in between the defects makes the geometry expand.  By adjusting the radii of the rings we can make these two effects balance out, and then add another concentric ring, and another, ad infinitum.  Figure~\ref{fig:DoubleWedge} shows the first two circular defect arrays (on the surface of time reflection symmetry) in such a ``bag of gold'' geometry (in this case with $p=3$ and $\sfk=2$).  We have tiled the Poincar\'e disk with hyperbolic triangles to make the hyperbolic geometry easier to visualize.  Because all hyperbolic triangles in the figure are isomorphic, both the solid red and dashed red lines are closed geodesics of the same length; the two curvature effects are precisely in balance.  The tubular region between the dashed and solid red geodesics (containing a ring of three defects) is a ``bag of gold'' building block $\cB$.    We can then cut the geometry apart along the outermost (solid red) geodesic, and insert as many copies of $\cB$ as we want, glued end to end, to make the bag of gold depicted in figure~\ref{fig:BagofGold}.  Again the location of each defect is a modulus which can be varied over some domain within the ``bag of gold''.  As we vary the locations of the defects, the lengths of the closed geodesics (and hence the areas of the various apparent horizons) increase or decrease in some range.

Clearly there is an infinite variety of such bag-of-gold geometries that can be constructed, by varying the type, number and placement of the defects.  A variant of the above construction has been employed in~\rcite{Balasubramanian:2022gmo,Balasubramanian:2022lnw} to argue that such states can be taken as a coherent state basis for black hole microstates.  As mentioned above, these configurations are extremely atypical, the equivalent of using a basis of microstates of a gas in a box where one has put all the molecules into a tiny corner of their phase space, and then relied on the fact of their thermalization to claim that one has captured the generic microstate.  Classically, this may be true, because of the infinite precision with which one can specify classical trajectories.  Quantum mechanically, working in a tiny corner of phase space is not sampling the generic microstate.  The approach of~\rcite{Balasubramanian:2022gmo,Balasubramanian:2022lnw} is to produce enough such outlandish configurations that one may argue that one has indeed sampled the phase space.

Perhaps it is more accurate to say that one has sampled the phase space of geometrical configurations on the surface of time reflection symmetry (for non-rotating black holes).  All these configurations evolve rapidly (in a proper time of order the $AdS$ scale $\ell$) to a stringy regime where the singularity of classical general relativity is expected to be resolved by strongly-coupled string theory.  A more accurate picture of the quantum wavefunction might therefore be that the wavefunction has two regimes or branches, that one might label geometric and non-geometric (or in the terminology of~\rcite{Martinec:2019wzw}, following~\rcite{Bena:2012hf,Lee:2012sc,Martinec:2015pfa}, ``Coulomb'' and ``Higgs''), and that the geometrical construction of~\rcite{Balasubramanian:2022gmo,Balasubramanian:2022lnw} captures the geometrical or ``Coulomb'' branch.  Something similar is seen in the Euclidean ``cigar'' geometry of the $\sltwo/\uone$ CFT that is the worldsheet description of Euclidean black fivebranes.  There the background is that of a geometrical cigar with metric
\be
ds^2 = n_5\lstr^2\big( d\rho^2 + \tanh^2\!\rho \, d\tauE^2 \big)
\ee
that is inextricably linked to a string winding condensate
\be
\langle \cV \rangle \propto \exp\big[-n_5(\rho+i\tilde\tauE)\big]
\ee
that was interpreted in~\rcite{Giveon:2015cma,Giveon:2016dxe} in terms of the vacuum string wavefunction again having two branches.  Moreover, the winding condensate of the Euclidean theory has been suggested (see \eg~\rcite{Kutasov:2000jp}) to analytically continue to a Hagedorn state of fundamental strings near the horizon in Lorentz signature (since the local Unruh temperature a string length from the horizon is the string scale).
It might be that a similar structure characterizes the non-perturbative wavefunction of the BTZ black hole in string theory~-- that there is a geometrical branch of the wavefunction that includes all the ``bag-of-gold'' geometries described above, and a stringy branch supported on Hagedorn configurations of the little string that opens up near the singularity of the classical geometry, but also extends out to the horizon scale in order to resolve the information paradox.  The perturbative string condensate seen in the above coset is also a property of the parent worldsheet theory on the Euclidean BTZ background $\hthree/\bZ$, with the same candidate interpretation.  One might say that the Hagedorn structure of the perturbative string near the horizon is simply the ``Coulomb branch'' tail of the Hagedorn little string wavefunction that governs the ``Higgs branch'' and the statistical physics underlying BTZ black hole thermodynamics.

While we have described the Euclidean evolution as preparing a state at the surface of time symmetry, that then further evolves as a black hole and develops a singularity of the effective geometry, it is perhaps more accurate to say that the singularity is already there in the state, and that the ``Coulomb'' and ``Higgs'' aspects of the wavefunction are both present.  There is no invariant bulk notion of time, and therefore of spatial slicing of the geometry; any surface anchored to a given time on the conformal boundary is equivalent, due to Hamiltonian constraints in the bulk gravity theory.  In any invariant characterization of the state, the wavefunction will contain both the black hole geometry, and the resolution of its singularity via a Hagedorn gas of little strings.

The existence of a pristine geometrical black hole interior, largely decoupled from the little string dynamics, cannot persist for long after the black hole forms; very likely it is a transient phenomenon~-- otherwise one has the Hawking process going on at the horizon, and the attendant paradoxes.  Indeed, unless there are approximate conservation laws that sequester the geometrical collective modes from the underlying chaos of black hole dynamics, one expects these modes to mix strongly with the little string degrees of freedom on time scales of order the scrambling time $\tau_{\rm scr}^{~} \sim \frac{\beta}{2\pi}\log({S-S_0})$.  After that, any geometrical picture of the interior should receive substantial modifications.

How then should we think of the flow of time, and causal relations in the black hole interior?  This is crucial for the resolution of the information paradox.  One thing that string theory has, that general relativity does not, is the underlying fivebrane dynamics, about which the perturbative string is giving us substantial evidence.  Perturbative strings see a causal structure which is that of the effective geometry with its horizon as an effective causal barrier.  If string theory resolves the information paradox, it may be because there is a separate clock provided by the underlying brane dynamics, and evolution in that clock maintains coherence of the wavefunction between the singularity and the horizon.  

For instance, the redshift to the horizon of a near-extremal black hole is not infinite, as the classical geometry suggests; rather, there is a finite depth to the throat, and a finite redshift, after quantum fluctuations of the geometry are taken into account~\rcite{Heydeman:2020hhw,Lin:2022rzw,Lin:2022zxd}.  Quantization of the modes captured by JT gravity results in a gap in the spectrum of order $1/N$ between the BPS ground states and the lowest excited states, indicating a maximum redshift in the geometry consistent with the weak-coupling, Hagedorn little string picture of the state space.  This result indeed suggests the absence of a causal horizon, and an IR dynamics governed by the internal clock of the fivebranes.

In this picture of the black hole, objects in the geometrical regime are swept into the interior where they fractionate into little strings.  The Hagedorn little string gas, coherent at the horizon scale, boils off Hawking quanta that unitarily radiate information back out into the geometrical regime.  Outside the horizon, the little strings are effectively confined; only the singlet modes~-- fundamental strings~-- are light excitations here, and one has the conventional geometry and dynamics of the black hole exterior.

The above scenario shares with recently proposed ``wormhole dynamics'' scenarios, or ``ER=EPR'' proposals (see~\rcite{Almheiri:2020cfm} for a review), the idea that there are alternative pathways for information flow, beyond what one sees in the classical black hole geometry.  The difference here is that there is no role for the conventional process of pair creation of Hawking quanta at the horizon.  Instead, the Hawking quanta emerge at the horizon from the Hagedorn gas of little strings.  

A particularly vivid example of this phenomenon is provided by the Hawking radiation of wound strings from the BTZ black hole~\rcite{Martinec:2023plo}.  Here, one is radiating strings which contribute to the background F1 charge, drawing them from the electric $H_3$ flux carried by the fivebranes.  In the black hole phase, the string winding charge is the winding charge of little strings (each unit of fundamental string winding fragments into $n_5$ little string windings, so that the total winding of the latter is $N=n_1n_5$).  A little string gas that correctly accounts for the entropy will also correctly account for the radiation rate since, as shown in~\rcite{Martinec:2023plo,Martinec:2023iaf}, the emission probability is entirely governed by the available phase space.

Unitarity of the evaporation process requires that the string being radiated be reconstituted from the underlying little string gas; but that string cannot have tunneled to the horizon from the singularity of the classical BTZ geometry; that gives the wrong amplitude, and thus the wrong thermodynamics.  The radiated string also cannot come from the Hawking process, which generates a negative charge cloud near the horizon from vacuum pair creation, that would be detected by the Gauss law, and leaves unmodified the original charge near the singularity, also detected by the Gauss law.  A geometrical picture of the interior would have these two charge distributions well-separated from one another, leading to the standard information paradox problem of the radiated strings carrying away no information about the initial black hole state.   

In the version of the fuzzball scenario presented here, the string emission instead proceeds via $n_5$ little string windings coalescing from the little string branch of the wavefunction and depositing a fundamental string onto the geometrical branch of the wavefunction.  


\vspace{1cm}

\section*{Acknowledgements}

I thank 
Samir Mathur 
and 
David Turton for discussions.
The work of EJM is supported in part by DOE grant DE-SC0009924.



\newpage

\bibliographystyle{JHEP}      

\bibliography{fivebranes}


\end{document}